\documentclass[10pt,a4paper]{article}

\usepackage[utf8]{inputenc}
\usepackage[T1]{fontenc}
\usepackage{microtype}
\usepackage[sc]{mathpazo}
\usepackage[hyphens]{url}
\usepackage{hyperref}

\usepackage{graphicx}

\usepackage[bottom, hang]{footmisc}
\usepackage[
    top=2.5cm,
    bottom=2.5cm,
    left=2cm,
    right=2cm,
    footskip=1cm,
    headsep=0.75cm,
    columnsep=20pt,
]{geometry}

\usepackage{fancyhdr}
\patchcmd{\maketitle}{plain}{empty}{}{}

\usepackage{natbib}
\usepackage[dvipsnames]{xcolor}
\setcitestyle{authoryear}
\usepackage{titling}
\pretitle{\begin{center}\huge\bfseries}
\posttitle{\end{center}}
\patchcmd{\maketitle}{plain}{empty}{}{}

\usepackage{booktabs}

\usepackage{caption}
\usepackage{enumitem}
\setlist{noitemsep}

\usepackage{hyperref}

\usepackage{amsmath,amssymb}

\title{Talking Surveys: How Photorealistic Embodied Conversational Agents Shape Response Quality, Engagement, and Satisfaction}

\author{Matus Krajcovic\textsuperscript{1,2} , Peter Demcak\textsuperscript{1} and Eduard Kuric\textsuperscript{1,2}\thanks{Corresponding author: \href{mailto:eduard.kuric@stuba.sk}{eduard.kuric@stuba.sk}\\ORCID(s): 0000-0001-9030-7337 (M. Krajcovic), 0000-0002-4111-1052 (P. Demcak), 0000-0002-7371-5512 (E. Kuric)\\ \textbf{Citation:} Krajcovic M., Demcak P., Kuric E. (2025) Talking Surveys: How Photorealistic Embodied Conversational Agents Shape Response Quality, Engagement, and Satisfaction\\}
}

\date{
\footnotesize\textsuperscript{1}
Faculty of Informatics and Information Technologies, Slovak University of Technology, Ilkovicova 2, Bratislava, 84216, Slovakia\\ \textsuperscript{2}
UXtweak Research, Cajakova 18, Bratislava, 81105, Slovakia\\
}

\begin{document}

\maketitle

\begin{center}
\large{\textcolor{red}{This is an original manuscript of an article published by Springer Nature in Behavior Research Methods on 29 June 2026, available online:} \url{https://doi.org/10.3758/s13428-026-03091-0}}
\end{center}

\begin{center}
\normalfont\bfseries\vspace{0.5\baselineskip} \abstractname
\end{center}
\begin{quote}
\normalfont\small
Embodied conversational agents (ECAs) are increasingly more realistic and capable of dynamic conversations. In online surveys, anthropomorphic agents could help address issues like careless responding and satisficing, which originate from the lack of personal engagement and perceived accountability. However, there is a lack of understanding of how ECAs in user experience research may affect participant engagement, satisfaction, and the quality of responses. As a proof of concept, we propose an instrument that enables the incorporation of  conversations with a virtual avatar into surveys, using on AI-driven video generation, speech recognition, and Large Language Models. In our between-subjects study, 80 participants (UK, stratified random sample of general population) either talked to a voice-based agent with an animated video avatar, or interacted with a chatbot. Across surveys based on two self-reported psychometric tests, 2,265 conversation responses were obtained. Statistical comparison of results indicates that embodied agents can contribute significantly to more informative, detailed responses, as well as higher yet more time-efficient engagement. Furthermore, qualitative analysis provides valuable insights for causes of no significant change to satisfaction, linked to personal preferences, turn-taking delays and Uncanny Valley reactions. These findings support the pursuit and development of new methods toward human-like agents for the transformation of online surveys into more natural interactions resembling in-person interviews.
\end{quote}

\begin{quote}
\end{quote}

\section{Introduction}

Unmoderated user research is on the rise, offering quick insights that empower rapid decisions and fine-tuning of user experience and design \citep{he2021}. Without a flexible and well-trained moderator overseeing the study’s administration, the responsibility falls to an automated system—the user interface of the tool, research protocol, prototype design and so forth—to ensure proper participant engagement. This is essential to collect useful data and avoid wasted research resources \citep{schirra2023}. Until now, the benefits typically gained from human interviewing—rapport-building, personal and personalized touch in collection of feedback, clearing up misunderstandings—were accepted as an inherent and inevitable tradeoff, with methods designed to partially mitigate its impact \citep{ward2018, kim2019, khayyatkhoshnevis2022}. However, more comprehensive solutions might be within reach.

Embodied conversational agents (ECAs) are intelligent software entities, designed for interactive verbal and nonverbal communication, visually represented as anthropomorphic characters \citep{provoost2017}. Driven by AI, they can act as mediators for digital communication \citep{hancock2020}. Recent advancements in technologies like natural language processing (NLP) and generative artificial intelligence (GAI) have enhanced their conversational ability by dynamically predicting appropriate responses \citep{lim2025, yang2025} and enabled creation of photorealistic embodiments \citep{tu2021}. However, there is currently a lack of validated instruments incorporating ECAs in this context. The readiness of ECAs to take on the role of a user research moderator remains an open question. The most closely related study was in the domain of healthcare, where \citet{zhu2025} investigated the effect of ECAs on collection of sensitive information (e.g., tobacco use, sexual behaviors) rather than explorative information gathering in user research. Furthermore, they were limited to static rule-based interactions (no conversational AI), and most participants were students.

Aiming to address the research gap, the chief contributions of our research include:
\begin{itemize}
    \item Proposal and technical implementation of Virtual Agent Interviewer, a prototype instrument for conducting surveys mediated by an Embodied Conversational Agent (ECA), utilizing current AI-driven technologies like Large Language Models (LLMs).
    \item Between-subjects experiment with 2 agents (embodied and chatbot) employed across 2 questionnaire surveys, yielding 2,265 responses from follow-up conversations by 80 participants sampled from the general population in the UK.
    \item Comprehensive mixed-methods evaluation of response quality, user engagement and satisfaction, supporting the positive impact of ECAs for surveys and providing insights for tuning future solutions. 
\end{itemize}

The following sections contain (\ref{sec:background}) the background, analyzing the current knowledge and the state of the issue, (\ref{sec:design}) our solution and the proposal of our instrument, (\ref{sec:method}) the experimental method for the instrument’s validation, (\ref{sec:results}) the result data and findings, (\ref{sec:discussion}) the discussion of implications, limitations and subjects of interest for future study, and finally (\ref{sec:conclusion}) the conclusion.

\section{Background}
\label{sec:background}

Prior studies have explored the potential of AI and machine learning to enhance collection of data in user research, such as by incorporating LLM chatbots to improve interactivity \citep{kim2019, xiao2020} or using deep-learning-based eye tracking to extract additional information from webcam footage \citep{kuric2025et}. However, applications of AI-driven anthropomorphic agents that leverage multiple types of AI models at once to tackle complex problems of unmoderated research remain scant. This section elaborates on the status quo, analyzing the problem along with an overview of relevant studies and theory.

\subsection{Moderation in user research}
In participant-based research, moderation is a means for researchers to flexibly facilitate collection of data \citep{bordegoni2023}. Trained moderators can recognize appropriate situations to interject \citep{debleecker2018}, enabling them to provide additional instructions or ask ad-hoc questions. Conversations between participants and moderators are dynamic, with rotating roles of the speaker and the listener. Organic conversations can be supported by various verbalizations, including those of seemingly minimal semantic value like affirmations (e.g., “okay”, “mm hm”) \citep{hertzum2018}. The flexibility of the moderator and natural conversations allows for adaptive solving of issues, such as confusing directions, or a prototype not displaying as intended on the participant's device \citep{schirra2023}. The accessibility of research can thus be improved for persons who may otherwise find it challenging, like people with cognitive impairments \citep{kleban2024}.

Despite the many advantages of moderated user research, unmoderated research has grown progressively more popular over the period of the last two decades, with crowdsourcing of unmoderated user studies now being an established industry \citep{schirra2023, liu2013, kittur2008}. Its advantages—namely remote administration, scalability through crowdsourcing from large pools, and cost-effectiveness—can be traced back to the increased prominence of web-based research as a de-facto go-to approach substituting previous paper-based methods \citep{hohwu2013}, further amplified by the rising number of internet users \citep{huang2006}. The ecological validity of studies can also be kept intact in the realistic conditions of the participant’s home. The absence of moderation can impact the feedback given by participants, such as by increasing their ratio of relevant verbalizations, since fewer verbalizations are process-related \citep{hertzum2015}.

While moderated and unmoderated research are both subject to the evaluator effect to a similar degree \citep{hertzum2013}, unmoderated usability testing can yield less reliable data that needs to be filtered for quality \citep{khayyatkhoshnevis2022}. The lack of user engagement is linked to careless responding \citep{ward2018} and satisficing (responding in passable but superficial, cognitively passive manner) \citep{kim2019} as sources of low-quality responses. Limited modality of communication can increase attrition, reducing the statistical power of surveys, as well as their internal validity, given that drop-off can correlate with variables such as personality traits \citep{ward2017}. These issues underscore the potential for augmenting unmoderated research through the development of methods that enhance interactiveness and feel more human.

\subsection{Conversational agents}
Colloquilally referred to as chatbots, conversational agents are technologies capable of communicating with users in natural language. Whether rule-based or AI-driven, their central challenge resides in natural language processing aimed at generating appropriate responses to user inputs \citep{khan2017, ren2022}. Large Language Models (LLMs) like ChatGPT brought significant advancements in the field, as corroborated by their good performance in language tasks \citep{iizuka2023}. Their advent contributed to significant expansion of chatbot usage in various fields, including e-commerce, finance, travel and gaming \citep{bilquise2022, yenduri2024}.

Conversational agents can augment user research, like traditional surveys, by transforming the process into social interaction \citep{celino2020, liu2024}. In choice-based surveys, \citet{kim2019} demonstrated the concept by incorporating LLMs and comparing four conditions with two independent variables: standard survey vs. chatbot and formal vs. casual communication style. The chatbot with casual style reduced satisficing and contributed to higher enjoyment. \citet{xiao2020} expanded the concept to chatbots asking open-ended questions. Analysis of textual responses based on Gricean maxims as indicators of communication quality corroborates that chatbot surveys are preferable for collecting high-quality information. Nonetheless, introduction of AI chatbots into research interactions can pose new challenges. \citet{kuric2024gpt} assessed the capability of LLMs to ask follow-up questions in usability testing. While providing more detailed information, repetition triggered frustration and led to less relevant responses.

In spite of fostering engagement, text-based conversational agents are still limited to a single channel of communication. Anthropomorphized agents could tackle the risks of detachment and disengagement common for unmoderated online research by evoking in participants a perception of proximity to the researchers and raising their sense of accountability \citep{ward2017}.

\subsection{Embodied conversational agents and virtual avatars}
By combining multiple modalities of humanlike appearance and behavior (speech, animated face with lip syncing, gestures, facial expressions), embodied conversational agents (ECAs) can create the impression of socially interacting with a person in possession of a physical body \citep{kipp2010, lugrin2022}. The term virtual avatar denotes embodied representations of agents in virtual environments, but also of users, who may hold a sense of embodiment towards their avatar \citep{genay2021, kim2025}. Since simulation of a realistic person is highly complex, most research of ECAs is task-oriented or specialized, originating from domains where human contact is critical for building rapport and engagement. Prevailing domains include healthcare, psychology and education \citep{yang2025, terstal2020, provoost2017}. Other use cases include marketing, e-commerce, customer support and gaming \citep{yang2025, qu2025}.

Realism and anthropomorphism are central concepts for ECAs. Increasing realism is a common goal that can improve the quality of user-agent interactions \citep{groom2009, qu2025}. Depending on its implementation however, high realism can sometimes be to a detriment. Evidence supports the Uncanny Valley theory, a concept originally from robotics, which states that small imperfections of entities bearing strong human likeness (e.g., virtual agents, humanoid robots) can evoke more disturbing and eerie experience in humans than entities designed to resemble humans only moderately \citep{groom2009, seyama2007, reinhardt2020}. Preferences toward the degree of realism can vary between individuals \citep{tan2025}. User behavior can also be affected by the agent’s appearance and familiarity, as \citet{tan2025} demonstrated with students in an educational system.

Inclusion of embodied agents with human faces in user interfaces tends to result in more positive interactions and attitudes \citep{yee2007}. A visible agent has the advantage of additional communication cues, such as eye contact, which can contribute to more favorable perception of pragmatic aspects such as ease of use \citep{reinhardt2020}. Effective nonverbal communication by ECAs is an ongoing subject of study, with gesture generation being a prominent aspect alongside the likes of facial expressions and posture, which can impact the understanding of communicated information \citep{wolfert2022}.

Investigations of ECAs in the context of user research are limited, with related studies primarily dedicated to assessment of mental health and diagnostics. \citet{auriacombe2018} and \citet{bickmore2020} conducted studies to validate agents for diagnosis of alcohol, tobacco and drug use disorders. In the most closely related study aimed at ECAs in psychosocial assessment, humanlike agents were found to encourage self-disclosure but discourage sharing of sensitive health information \citep{zhu2025}. However, (a) it was focused primarily on sensitive health questions rather than making user research more interactive, (b) did not incorporate LLMs or other AI models, (c) was restricted to a sample of young adults (students) and (d) limited its analysis of engagement to word count, reducing its findings about the effects of ECAs on the open-ended responses that are applicable for user research. Furthermore, advancements in AI make virtual avatars with more realistic appearances possible.

Currently, ECAs are undergoing an AI wave \citep{schobel2024}. Fueled by advances in generative artificial intelligence (GAI) and natural language processing (NLP), they aim to overcome some of their past limitations. Aside from leading speech-based conversations, AI-based tools such as HeyGen\footnote{HeyGen AI video generator: \url{https://www.heygen.com/}}, Synthesia\footnote{Synthesia AI video platform: \url{https://www.synthesia.io/}} or Azure AI text to speech\footnote{Azure text to speech avatar: \url{https://learn.microsoft.com/en-us/azure/ai-services/speech-service/text-to-speech-avatar/what-is-text-to-speech-avatar}} can generate videos driven by data, resulting in more realistic appearances than 3D models rendered by standard computer graphics (such as those investigated by \citet{zhu2025}). The need to manually model, rig and animate 3D assets is eliminated. Although proprietary, these solutions feature pre-built customizable models, facilitating their adaptation for broad use cases. Examples of their application include generation of educational videos for students \citep{contreras2024} or personalized delivery of information to medicinal patients \citep{badawy2025}.

The advent of aforementioned AI-based video technologies sets the stage for innovation in user research. As this article aims to demonstrate, virtual agents simulating aspects of a human moderator could support interaction with participants through added modalities and positive engagement. This may allow unmoderated user experiments to reap some of the benefits traditionally associated with moderation, such as higher quality of data through reduction of satisficing and careless responding.

\section{Instrument design}
\label{sec:design}

To create more engaging and natural-feeling interactions in user research, we propose simulated moderation that leverages the strengths of human-like ECAs integrated with AI-driven conversational models. As a paradigm commonly embraced for its convenience and scalability, unmoderated research holds great potential for an enhancement that could mitigate its inherent challenges: the lack of human contact and spontaneous interactivity. This pursuit aligns with goals of human-centered artificial intelligence, promoting augmentation over automation to extend the abilities of researchers \citep{esposito2024}.

With that overarching ambition, this research was conducted as a proof of concept, balancing its vision with a manageable scope. Moderation in user research, when performed by humans, is a skill with many facets, requiring the ability to flexibly react to unique emergent problems, to know when and how to intercept, how to ask questions and spot contradictions between expectations, participant statements and behavior \citep{kuric2024gpt}. These are complex capabilities that would merit individual research. Therefore, the prototype introduced in this section is designed to isolate and assess the effects of ECAs on user research in their base form.

\subsection{Prototype design decisions}
We introduce the Virtual Agent Interviewer (VAI), a prototype of an online survey tool that incorporates an AI-driven ECA into its survey flow. This agent can perform certain functions of a moderator by interacting with the participants, asking them follow-up questions to obtain deeper insights about answers. Additionally, it can facilitate messaging of instructions and debrief sessions by interactively answering questions from participants. The following decisions were incorporated in its design and implementation:
\begin{itemize}
    \item \textit{Embodiment as a feature of focus.} Construction ECA-enabled moderation could encompass a variety of features and subsystems. However, the first priority is to assess whether the embodiment of the agent (face, gestures, speech modality) by itself contributes value, and no significant issues are involved with its application.
    \item \textit{Survey augmentation.} Unmoderated user research subsumes a variety of techniques that can  differ significantly by their mode of interaction. Because of this diversity, surveys were chosen as an appropriate baseline technique for the adoption of the Virtual Agent Interviewer. Surveys are commonly conducted online, without moderation, and they can be procedurally simplistic in comparison to more involved techniques such as usability testing. Furthermore, asking follow-up questions in surveys represents an intuitive application for the Virtual Agent Interviewer. Aside from the ability of the AI model to generate meaningful questions, no additional intelligent components are required (e.g., automated detection and interpretation of usability issue encounters to moderate usability testing). 
    \item \textit{Text-based baseline.} The VAI can be viewed as a natural progression from surveys augmented with AI chatbots \citep{xiao2020, kim2019}, i.e., disembodied text-based agents. Therefore, a text-based conversational agent was implemented in the survey tool as a baseline alongside the embodied agent, ensuring comparison under equal conditions.
    \item \textit{Custom survey tool.} To include the embodied and text-based conversational agents in surveys, we developed a new generic web-based tool rather than integrate agents into existing tools. The evaluation of the instrument is thus also isolated from latent confounding variables linked to the design of existing tools, potentially mitigating biases.
\end{itemize}

To generate photorealistic virtual avatars capable of leading quasi-real-time conversations, VAI uses the AI-driven solution Heygen Interactive Avatar\footnote{Heygen Interactive Avatar: \url{https://www.heygen.com/interactive-avatar}}. HeyGen was selected over its competitors for providing a broad selection of pre-built virtual avatars that were assessed as having more natural behavior and appearance (triggering lesser Uncanny Valley response) during the review of available technologies at the time of design and experiment planning (November 2024). For further technical details, see \autoref{app:implementation} \hyperref[app:implementation]{Technical implementation}

\subsection{Survey structure and functionality}
\label{sec:structure}

\begin{figure}[!ht]
    \centering
    \includegraphics[width=0.8\linewidth]{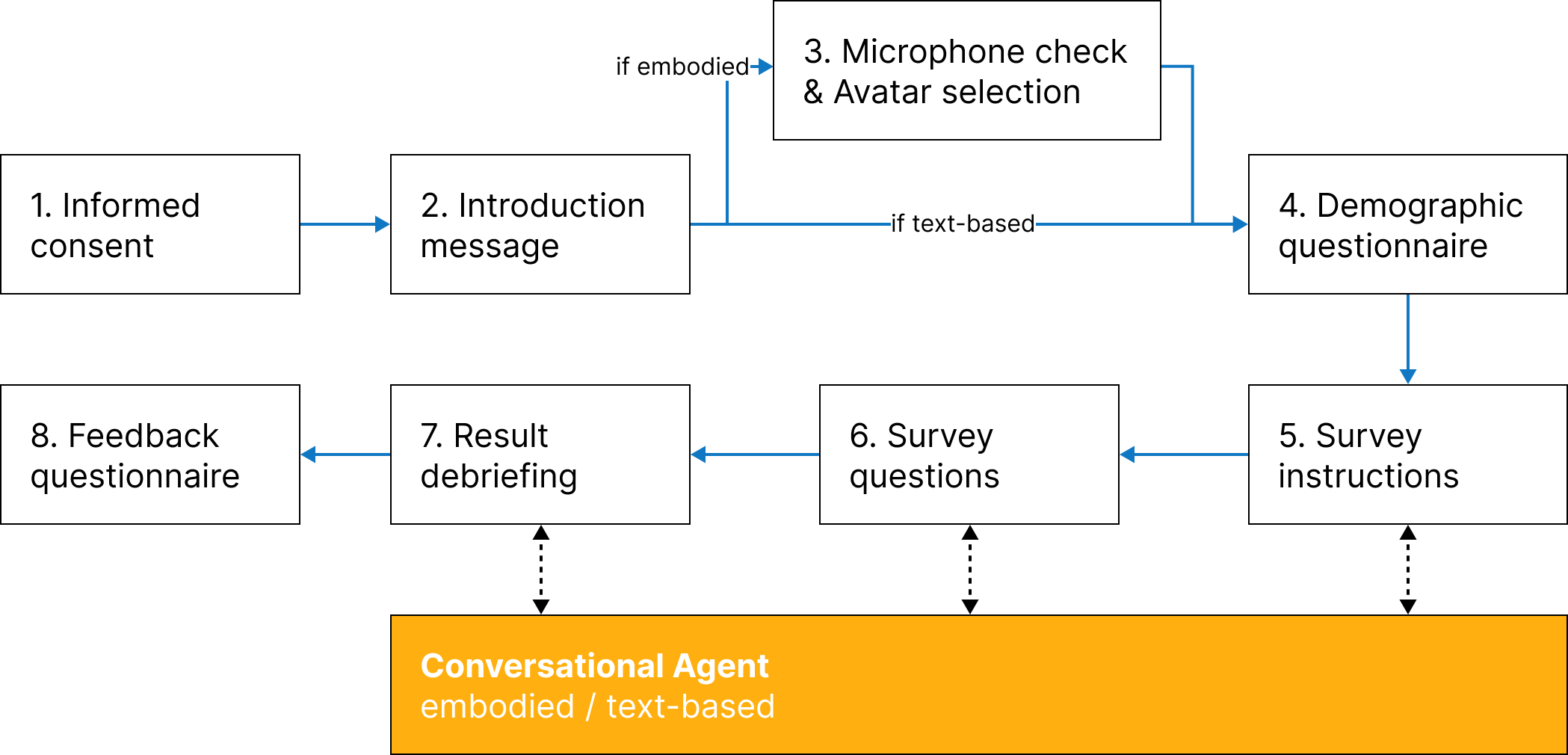}
    \caption{Overview of the VAI survey flow template.}
    \label{fig:survey-flow}
\end{figure}

The VAI tool supports creation of surveys in two variants, either with the embodied conversational agent (face and speech), or the text-based conversational agent (chatbot). Surveys are structured in accordance with typical logical flow of online surveys and in compliance with ethical and user research standards. Their template is shown in \autoref{fig:survey-flow}.

Before any further steps, participants are first asked for their informed consent to collect data for research purposes. This is followed by a static introduction message where participants can be informed about what they can expect and what will be required of them. Because interaction with the ECA relies on the participant speaking into a microphone, this variant includes a microphone check to ensure validity of data. Participants are asked to enable their microphone and speak into it. Their speech is played back to them so they can verify that the microphone is functioning correctly.

\begin{figure}[!ht]
    \centering
    \includegraphics[width=\linewidth]{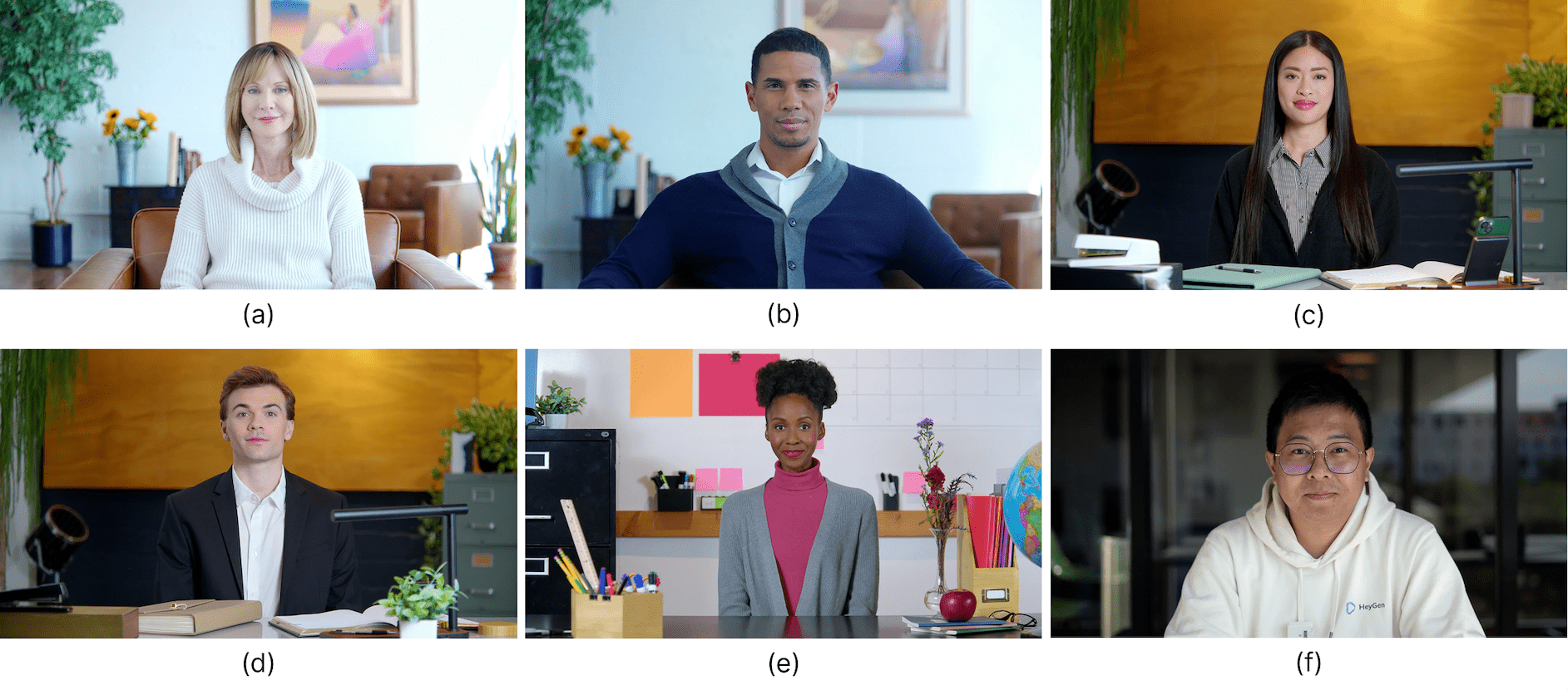}
    \caption{Avatar appearances that participants can choose between at the beginning of VAI survey ECA variant. Model names as adopted from Heygen are (a) Ann, (b) Shawn, (c) June, (d) Silas, (e) Judy, and (f) Wayne.}
    \label{fig:avatar-appearances}
\end{figure}

Additionally, to mitigate potential biases that participants may have toward the specific appearances of the ECA, participants in this variant are also asked to choose the appearance of the agent’s virtual avatar that they will be interacting with during the survey. Among the readily available appearances provided by Heygen, the six that are shown in \autoref{fig:avatar-appearances} are offered to participants. Given that attributes like race and gender can affect how users perceive and relate to agents \citep{zhao2023}, the criteria for selection of appearances were aimed at enhancing diversity of options. Each participant could thus pick an option they would find the most sympathetic.

Before the main body of the survey that entails interaction with the conversational agent, participants can be presented with a demographic questionnaire. This questionnaire is intended primarily to obtain descriptive data about the participants, such as gender, age, or education.

During the three segments of the surveys that involve interaction with conversational agents, the agent appears on the screen. Each conversation is initiated by the agent with a pre-defined conversation starter, such as a question, or a comment that encourages discussion of the current step of the survey.
\begin{itemize}
    \item The embodied conversational agent speaks to participants through voice. To maximize a feeling similar to a video call and having an authentic conversation with a person, the agent's avatar is always initially displayed using the full size of the browser window. The user can minimize the avatar to the bottom right corner afterward. Participants can reply after pressing a push-to-talk button. This reduces the chance of Heygen misinterpreting background noise as speech, while also giving participants an opportunity to think through their answers. When the user stops speaking for a few seconds, Heygen automatically stops the recording and proceeds with generating a reply. Replies can be up to 60 seconds long. A countdown is displayed after the first 45 seconds if the participant is still speaking, providing them time to finish naturally without sudden interruption.
    \item The text-based conversation agent is interacted with as a standard chatbot.
\end{itemize}

\begin{figure}[!ht]
    \centering
    \includegraphics[width=\linewidth]{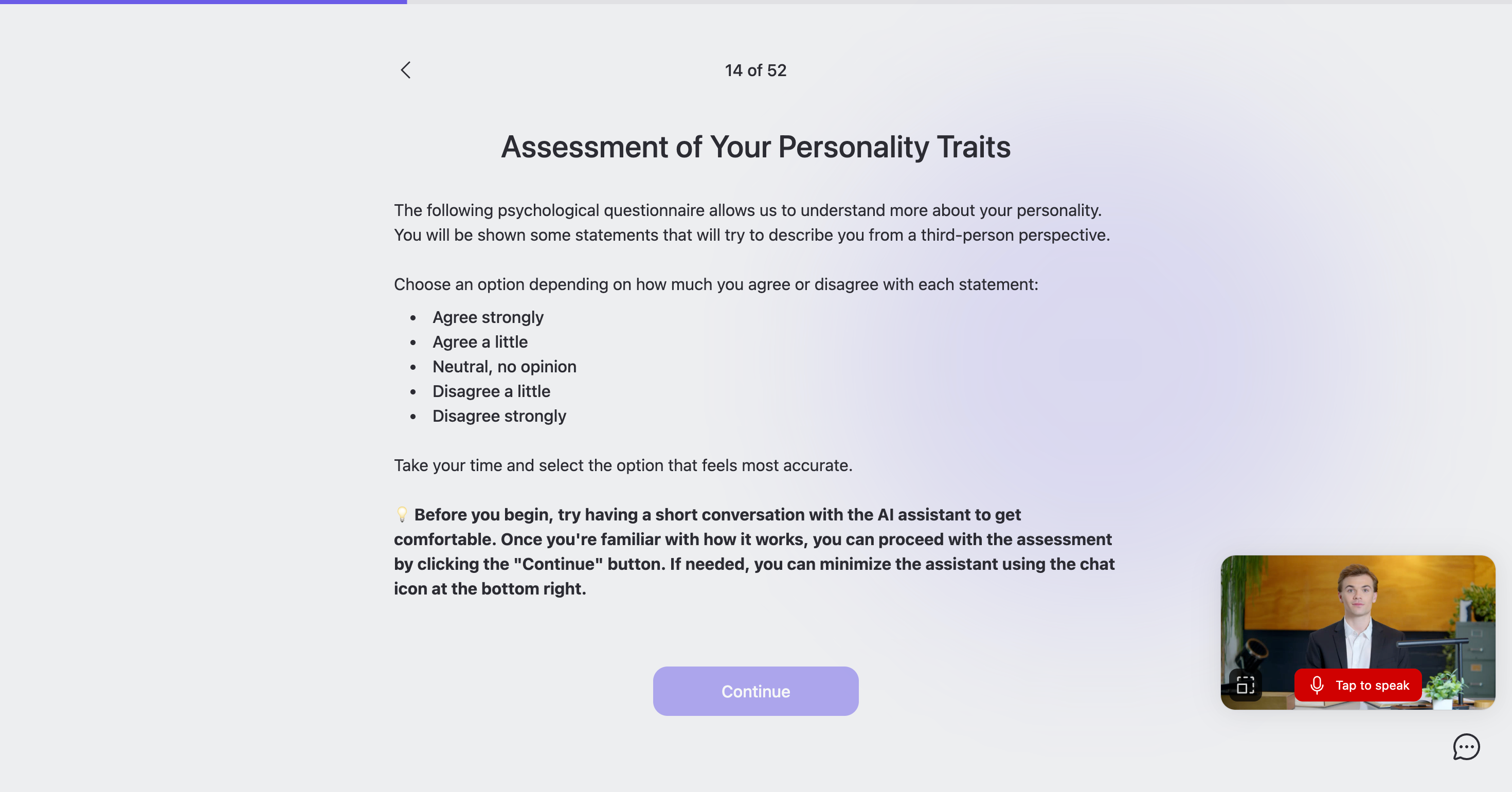}
    \caption{Survey instructions with the embodied conversation agent displayed in the bottom right corner after it has been minimized (it starts maximized to the full browser window). In text-based variant, the agent is displayed analogously to \autoref{fig:textbased-example}.}
    \label{fig:embodied-example}
\end{figure}

Survey instructions further specify what the survey is about and guide the participant on how to complete its questions. After reading the instructions, conversation with the agent is initiated by the press of a button. The conversation agent is included here as a warm-up as shown in \autoref{fig:embodied-example}, allowing participants to familiarize themselves with the agent, as well as interacting with it.

\begin{figure}[!ht]
    \centering
    \includegraphics[width=\linewidth]{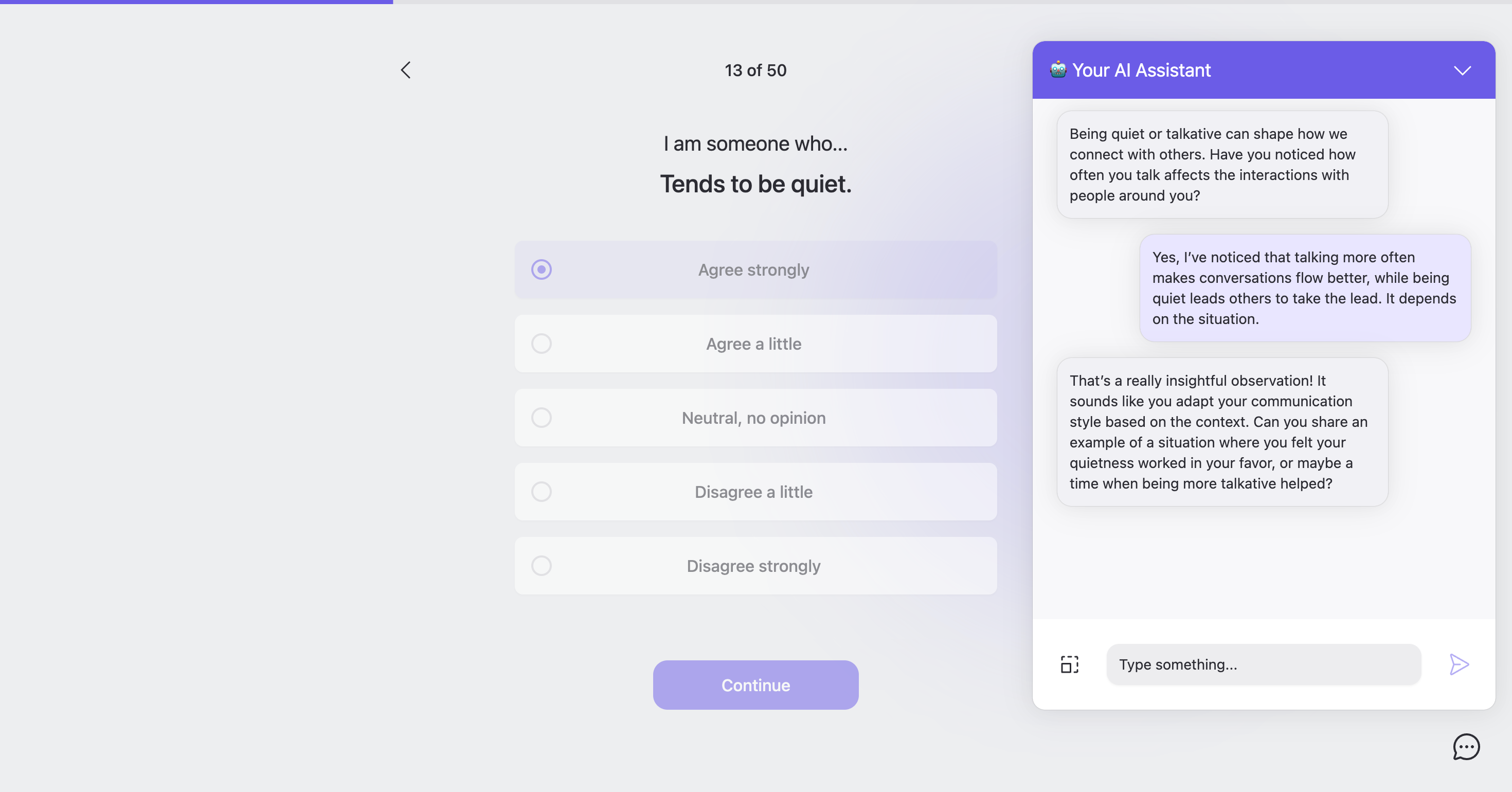}
    \caption{Survey questions with the text-based conversational agent follow-up popped out on the right side after an answer was submitted. In the embodied variant, the agent is displayed analogously to \autoref{fig:embodied-example}.}
    \label{fig:textbased-example}
\end{figure}

Survey questions are displayed one at a time as seen in \autoref{fig:textbased-example}. The conversational agent avatar or chatbot is displayed after the user submits their answer. The use of the agent in questions is optional, individual questions can be included in a questionnaire either with or without conversational follow-up.

Result debriefing is a segment where participants can review the information that has been collected. This can serve as a summary, or provide automated evaluation to facilitate self-discovery in case of questionnaires where results have intrinsic value to participants themselves (e.g., personality assessment or other psychometric tests). The agent is incorporated to optionally assist participants with interpreting the meaning of their results in an interactive and engaging manner.

Further details on how the logic of the agents is orchestrated are shown in \autoref{app:modularization} \hyperref[app:modularization]{Modularization and logic}. Once the survey (and interaction with the agent) is concluded, the questionnaire segment at the end collects final feedback from participants.

\section{Experimental method}
\label{sec:method}
An experiment was conducted to validate proposed facilitation of surveys by embodied conversational agents (ECAs). For comparison against a disembodied text-driven baseline alternative, we adopted a chatbot to facilitate surveys under identical conditions.

\subsection{Research questions}
The primary objective of this research is to evaluate the contribution of the Virtual Agent Interviewer toward surveys that provide better experience to participants, and thus also better responses to conductors. To obtain additional insights, we also direct attention at the differences between facilitated questionnaires and their impact on agent-participant interactions. We ask the following research questions:
\begin{itemize}
    \item \textit{RQ1: Does facilitation of surveys with an embodied conversational agent (ECA) yield different quality of responses to a text-based agent?} The quality of responses is critical for obtaining reliable and valid findings from surveys, as well as other user research methods. Text-based survey interactions yield high-quality results from a multitude of perspectives, including their informativeness, relevance, specificity and clarity. By inducing even more natural conversations, we extrapolate that even better results might be obtained by incorporating an ECA. We expect that incorporation of an ECA will improve the quality of responses.

    \item \textit{RQ2: Does facilitation of surveys with an ECA affect user engagement?} Engagement in surveys can serve as a signal of participant understanding and of more thorough, natural responses \citep{hess2013}. Nonetheless, there is merit to assessing it independently from response quality to capture potential dissociations. Text-based chatbot surveys raise engagement through custom interactions in comparison to traditional surveys \citep{xiao2020}. Therefore, conversations with anthropomorphic agents can be expected to boost engagement even further through more natural interaction involving natural speech and a human face \citep{yee2007}.

    \item \textit{RQ3: Does facilitation of surveys with an ECA affect user satisfaction?} Satisfaction from a pleasant interaction can support the intrinsic motivation of participants taking part in a survey \citep{hossain2012}. Therefore, we aim to assess scales about the attitudes of participants toward the interaction with a virtual agent. We also analyze qualitative feedback to identify factors that influence user satisfaction.

    \item \textit{RQ4: Do the latent characteristics of different surveys lead to differences in the interaction with—and results obtained through—an ECA?} Due to the heterogeneity of user surveys, with different questionnaire logic and structure applied for varied purposes, behavioral patterns linked to interactions with virtual agents may vary. The resulting usefulness of virtual agents (e.g., for eliciting higher engagement or better results) may depend on the idiosyncracies of questionnaires. Cognitively demanding questions might make it more difficult for participants to think through and express complex answers. Therefore, to obtain theoretical findings with higher generalizability and explanatory ability, we aim to compare the effects of ECAs on different questionnaires.
\end{itemize}

\subsection{Participants}
\label{sec:participants}

Eighty (80) participants were recruited for a between-subject experiment aimed at mitigating carryover bias that could otherwise affect observations in within-subject design \citep{zhu2025}. The sample size was determined using an a priori power analysis in G*Power\footnote{G*Power statistical power analyses: \url{https://www.psychologie.hhu.de/arbeitsgruppen/allgemeine-psychologie-und-arbeitspsychologie/gpower}}. The analysis was based on a two-tailed Wilcoxon-Mann-Whitney test, with 80\% power, an alpha level of .05, and a Cohen’s \textit{d} of 0.65 (approximately equivalent to a medium effect size of \textit{r} = 0.3), indicating that a total of 80 participants would be sufficient. The experimental conditions were defined by two surveys (see \ref{sec:materials} \hyperref[sec:materials]{Materials}) completed with one of the two conversational agent variants (text-based and embodied). To address risks to internal validity inherited from between-subject design, participants were sampled from the general population of internet users through a random stratified process, aimed at obtaining equal gender distributions and age-wise representation of the population according to statistical data\footnote{Distribution of internet users worldwide: \sloppy \url{https://www.statista.com/statistics/272365/age-distribution-of-internet-users-worldwide/}}. 

\begin{figure}[!ht]
    \centering
    \includegraphics[width=\linewidth]{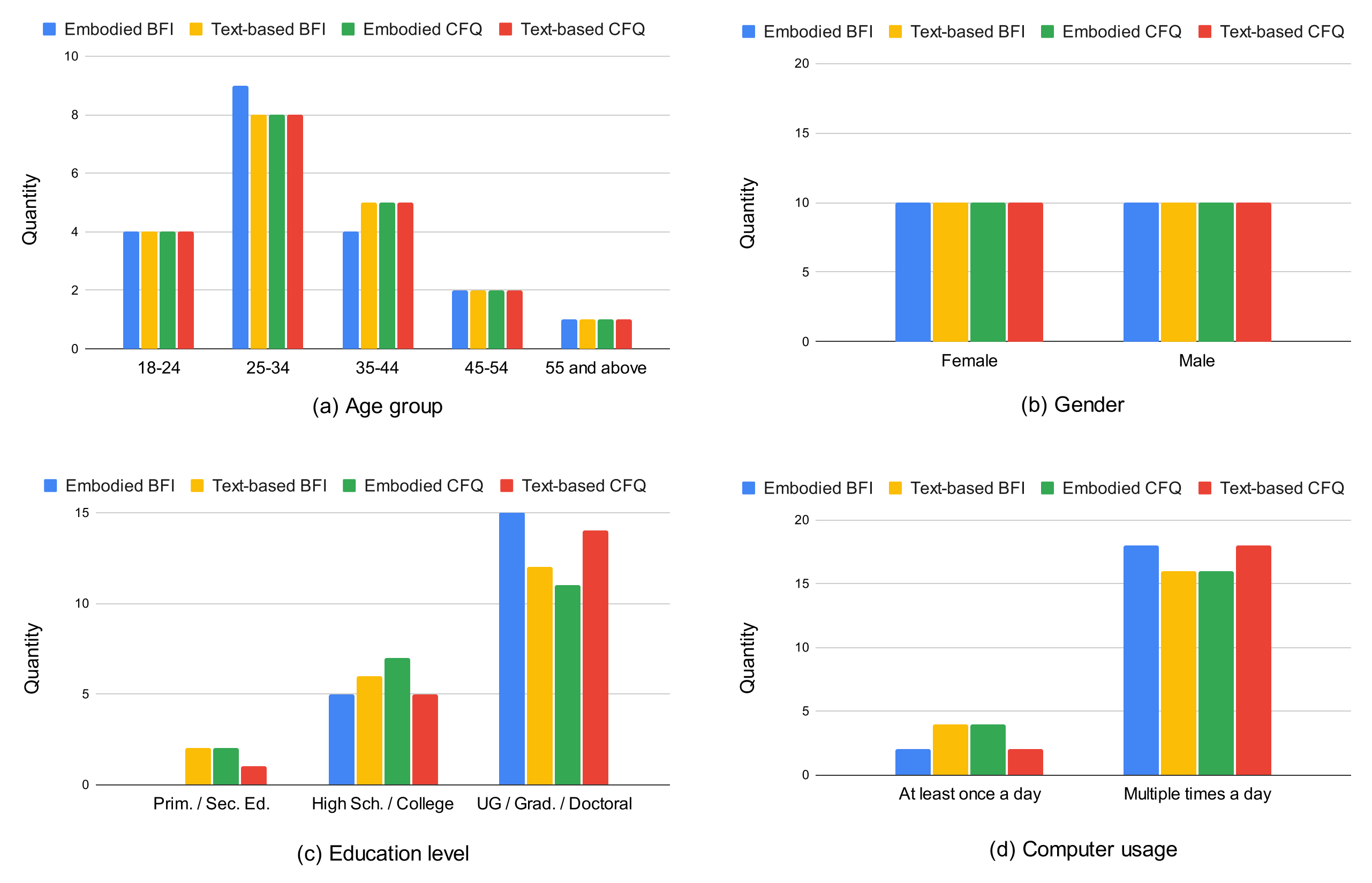}
    \caption{Participant demographics and computer usage—covering gender, age, education level, and frequency of computer use—were balanced across all study variants.}
    \label{fig:demographics}
\end{figure}

Participants were recruited in the UK through the online panel service of the UXtweak Research tool\footnote{UXtweak UX research tool: \url{https://www.uxtweak.com/}}. Participants were using desktop devices during the experiment. Descriptive attributes were balanced across conditions, including gender $(\chi^2(3,N=80)=0, p=1.0)$, age $(\chi^2(12,N=80)=0.25, p=1.0)$, education $(\chi^2(3,N=80)=3.45, p=.79)$, computer usage $(\chi^2(3,N=80)=1.57, p=.72)$ and attitude toward AI $(\chi^2(6,N=80)=1.27, p=.97)$, as shown in \autoref{fig:demographics}. Most participants expressed favorable or neutral stances toward AI: 17 rated their attitude as Very positive, 39 as Somewhat positive, 15 as Neutral and only 9 as Somewhat negative. The majority of participants (75 out of 80) had previous experience with a conversational agent (e.g., chatbot, virtual assistant, avatar).

\subsection{Materials}
\label{sec:materials}

Even small differences between surveys—such as the subject of the questions, the complexity of understanding them, and the difficulty of answering them—may influence the manner in which users interact with the embodied conversational agent. Therefore, we implemented two different surveys in VAI to enhance the external validity of its evaluation. By choosing standardized and well-recognized psychometric questionnaires where conversations might expand the understanding of participants, we aim to support the relevance as well as the reproducibility of our experiment. The following questionnaires were adopted into surveys:
\begin{itemize}
    \item \textit{Short Big Five Inventory 2 (BFI-2-S).} The BFI-2-S is a short version of the personality assessment questionnaire Big Five Inventory \citep{soto2017}, selected as a general questionnaire that most participants would be able to respond to without issue. It consists of 30 items for gauging personality traits on a 5-point scale that ranges from Strongly agree to Strongly disagree. In spite of its short format that reduces completion time, it also preserves much of the reliability of the full-length questionnaire. Each item corresponds to one of five personality trait scales: extraversion, agreeableness, conscientiousness, negative emotionality, and open-mindedness. The scales can be further divided into three facets.
    \item \textit{Cognitive failures questionnaire (CFQ).} The CFQ measures the likelihood of a person making cognitive mistakes and slips during everyday activities \citep{rast2008}. The reason for its selection is the higher sensitivity of its contents in comparison in BFI-2-S. Its inclusion aims to assess whether ECAs could build participant trust and aid in sharing such information. CFQ comprises 25 items (cognitive failures). Participants rate the frequency at which they undergo cognitive failures on a scale from Very often to Never. Higher scores suggest a greater tendency toward cognitive lapses. Three subdimensions of failures are present: forgetfulness, distractibility and false triggering (disruption when carrying out a series of mental or physical tasks).
\end{itemize}

The frameworks of the two questionnaires integrated the conversational agents to ask a series of open-ended follow-up questions. Since elaborating the answers to every question would be exhaustive for the participants, we consolidated the added interactivity with preserving survey conciseness. The follow-up was restricted to selection of items from both questionnaires (10 from BFI-2-S and 9 from CFQ). The rest of the questions remained without follow-up — a parallel to moderated interviews where selecting the topic for follow-up to focus on is commonly the prerogative of the moderator. The selection criteria for questions to receive a follow-up prioritized the representation of diverse scales and facets (personality traits/cognitive failures), questions with potential for meaningful in-depth conversations, and even temporal distribution throughout the questionnaire. Selected questions are summarized in \autoref{tab:followup-bfi} and \autoref{tab:followup-cfq}.

\begin{table}[!ht]
\caption{Questions with conversational agent follow-up in the BFI-2-S questionnaire. Numbers represent item order. Scales include E - extraversion, A - Agreeableness, C - conscientiousness, N - Negative emotionality, O - Open-mindedness.}
\label{tab:followup-bfi}
\begin{tabular*}{\textwidth}{@{\extracolsep\fill}llp{4cm}p{10cm}}
\toprule
\textbf{\#} & \textbf{Scale} & \textbf{BFI-2-S Question: I am someone who...} & \textbf{Initial follow-up} \\
\midrule
1 & E & Tends to be quiet. & Being quiet or talkative can shape how we connect with others. Have you noticed how often you talk affects the interactions with people around you? \\
5 & O & Is fascinated by art, music, or literature. & Art, music, and literature can really shape our lives in unique ways. How do they influence your day-to-day experiences or the way you see the world? \\
8 & C & Has difficulty getting started on tasks. & Getting started on things can be tough sometimes. Do you find certain types of tasks harder to begin than others? What helps you push through when that happens? \\
9 & N & Tends to feel depressed, blue. & What factors negatively influence your mood? Is there anything you do to help manage it? \\
11 & E & Is full of energy. & Having a lot of energy can shape your day in different ways. How do you usually channel your energy? Are there moments when it feels like too much or not enough? \\
14 & N & Is emotionally stable, not easily upset. & Is there anything you'd like to improve at, in regards to emotional stability? \\
17 & A & Can be cold and uncaring. & Have you noticed times when you feel more distant or detached? How do you think that impacts your relationships? \\
18 & C & Keeps things neat and tidy. & Do you usually stick to a set routine to stay organized, or do you prefer taking things one step at a time? \\
22 & A & Is respectful, treats others with respect. & Respect can mean different things to different people. What does treating others with respect look like to you in daily life? \\
30 & O & Has little creativity. & Can you name any situation that has allowed you to express yourself creatively and how? \\
\bottomrule
\end{tabular*}
\end{table}

\begin{table}[!ht]
\caption{Questions with conversational agent follow-up in the CFQ questionnaire. Numbers represent item order. Scales include F - Forgetfulness, D - Distractibility, FT - False Triggering.}\label{tab:followup-cfq}
\begin{tabular*}{\textwidth}{@{\extracolsep\fill}llp{5cm}p{9cm}}
\toprule
\textbf{\#} & \textbf{Scale} & \textbf{CFQ Question} & \textbf{Initial follow-up} \\
\midrule
2 & F, FT & Do you find you forget why you went from one part of the house to the other? & Can you recall any situations when your purpose for being in a room skipped your mind? Do you have any explanation for why it happened? \\
6 & FT & Do you forget whether you've turned off a light or a fire or locked the door? & Are there any specific switches or locks that you forget about, or does it happen in general. Do you have any little routines or reminders to help you? \\
7 & F & Do you fail to listen to people's names when you are meeting them? & Remembering names can be effortless for some and tricky for others. Are there people whose names you find easier or harder to remember? \\
9 & D & Do you fail to hear people speaking to you when you are doing something else? & Balancing attention between tasks and conversations can vary from person to person. How do you usually manage focus when you're engaged in something? \\
17 & F & Do you forget where you put something like a newspaper or a book? & Some people prefer to keep things in the same spot, while others leave them wherever they last used them. What approach works best for you, and why do you think that is? \\
19 & D & Do you daydream when you ought to be listening to something? & It's common for people to experience moments of distraction. Do you recall situations when it happens to you? \\
22 & F & Do you find you can't quite remember something although it's on the tip of your tongue? & Do you have any techniques to help you remember? How important are the things that you forget like this? \\
23 & F, FT & Do you forget what you came to the shops to buy? & When it comes to remembering what you need at the store, do you have a go-to trick or routine? Or do you just wing it and hope for the best? \\
25 & D & Do you find you can't think of anything to say? & Some conversations flow easily, while others can be more challenging. Do you notice any patterns in what makes a conversation feel easier or harder to engage in? \\
\bottomrule
\end{tabular*}
\end{table}

During an initial questionnaire, participants provided information about their age, gender, race, education, income bracket, English language proficiency, attitude toward AI, and level of previous experience with AI and conversational agents. The final questionnaire contained attitudinal Likert scales to obtain feedback about conversational assistants: 
\begin{itemize}
    \item \textit{How natural did the conversation with the AI assistant feel?}
    \item \textit{How would you rate the overall experience of interacting with the AI assistant?}
    \item \textit{How would you rate your overall experience with this survey?}
\end{itemize}

Additionally, participants were asked about whether they prefer the conversational agent that they interacted with, or the method of interacting with the alternative (text-based/embodied) agent, which was described to them. At the end, an optional open-ended question was used to give participants the opportunity to provide auxiliary feedback.

Attention check questions were included to monitor whether participants were putting genuine effort into reading and following instructions \citep{kuric2025usability, thomas2017}. To assess attention throughout the experiment, checks were incorporated in the initial questionnaire, in the middle of the main questionnaires (BFI-2-S or CFQ) and in the final questionnaire.

\subsection{Procedure}

To represent realistic online conditions that the VAI tool was designed for, the participants completed the survey remotely. The variability of conditions in the wild can be substantial due to an interplay of environmental and technological factors. Therefore, to alleviate concerns about ecological validity, the experiment was conducted in uncontrolled conditions that mirror the intended use of online surveys assisted by a conversational agent.

Participants, divided into four groups, each comprising 20 individuals (see \ref{sec:participants} \hyperref[sec:participants]{Participants}), were automatically assigned into corresponding experimental variants: E-BFI (embodied agent, BFI-2-S questionnaire), T-BFI (text-based agent, BFI-2-S questionnaire), E-CFQ (embodied agent, CFQ questionnaire), and T-CFQ (text-based agent, CFQ questionnaire). From the perspective of participants, the following procedure aligned with the structure of surveys in VAI as expounded in \ref{sec:structure} \hyperref[sec:structure]{Survey structure and functionality} and illustrated in \autoref{fig:survey-flow}. 

After participants confirmed their consent, an introduction message educated them about the core principles of the survey, including the concept of the conversational agent as an “interactive AI assistant”. Participants assigned to the embodied agent variants completed a microphone check and selected their preferred avatar (see \autoref{fig:avatar-appearances}). Through the initial questionnaire, they filled in their descriptive information. Survey instructions then guided participants on how to complete the Likert scales, depending on their assigned questionnaire variant (BFI-2-S or CFQ). Upon completing the main agent-assisted questionnaire, participants received a result debriefing. Personality trait and cognitive failure scores were calculated automatically and accompanied by basic definitions with examples. This allowed participants to either interpret their results by reading the static explanations, or through interactive discussion with the conversational agent. Lastly, summative feedback was collected through the final questionnaire, after which the participants were free to close the page.

\subsection{Data preprocessing and analysis}

Collected data was preprocessed to ensure its accurate analysis. Follow-up conversations contain 2,265 responses, after deducting 12 missing responses from the embodied agents and 3 from the text-based agents. The missing 0.7\% of responses were attributed to connection interruptions encountered as the study was conducted remotely, and to the instability of the virtual avatar (Heygen) API. The majority of data remained unaffected and could be analyzed as intended.

Responses in the text-based variant contained typos while transcripts from the embodied variant contained spelling errors and incorrectly joined words. To ensure correct tokenization for consistent evaluation of the agents, the SymSpell\footnote{Symspellpy spelling correction library: \url{https://symspellpy.readthedocs.io/en/latest}} library was employed to correct the format of user inputs. Examples of such corrections include:
\begin{itemize}
    \item Splitting words that were mistakenly joined together
    \begin{itemize}
        \item Mysupport → My support
        \item elderon a train → elder on a train
        \item helped meenjoy → helped me enjoy
        \item meet.At → meet at
    \end{itemize}
    \item Fixing typos
    \begin{itemize}
        \item dood;ing → doodling
        \item Irespectfully → respectfully
    \end{itemize}
\end{itemize}

Since the data distributions were non-normal, non-parametric Mann-Whitney U tests with effect size r were used to compare conditions across numerical and ordinal measures as well as the chi-squared test with effect size V for categorical measures. Each test involved a sample size of 80 (or 40), with one aggregated value per participant to maintain independence of observations assumed by the tests. Medians were analyzed to reduce the impact of outliers. For ordinal data (e.g., scales from 0 to 2), means were calculated for each participant.

\subsection{Measures}

The investigated dimensions of survey results and user experience are reflected in the analyzed measures. For continuity with prior work investigating text-based conversational agents in surveys by \citet{xiao2020}, we adopted theoretically consistent measures, further extending and adapting their set to enable a thorough and construct-valid comparison between agents.

Response quality. To tackle the challenge of assessing the quality of open-ended survey answers, we implement measures developed by \citet{xiao2020} based on the Grice’s maxims, four principles of effective communication. Calculated from the text of the participant’s responses in conversation with the agent, they include:
\begin{itemize}
    \item \textit{Informativeness.} Determined as a sum of the surprisal of words, which is the inverted frequency of a word in the English language. To determine frequency, we used the library wordfreq \citep{speer2022}, which collates text from various sources, including books, Wikipedia articles, news, social media posts and comments. 
    \item \textit{Specificity.} Defines the level of specific detail in the response that can help researchers obtain in-depth qualitative insights. Coded as integers on a range from 0 (general responses), through 1 (mentions of specific concepts), to 2 (concepts are further elaborated on through examples or other justifications).
    \item \textit{Relevance.} Conceptual match between the questions and the response, which can be warped due a variety of factors, such as misunderstanding or self-presentation biases. Coded as integers on a range from 0 (off-topic and nonsensical responses), through 1 (responses with indirect implications, but without providing a straight answer), to 2 (completely valid answers).
    \item \textit{Clarity.} The transparency of the meaning of the response, allowing for it to be interpreted and understood without ambiguity. Coded as integers on a range from 0 (illegible), through 1 (incomplete or partially legible, such as without proper sentences), to 2 (clear and well-articulated).
\end{itemize}

Engagement. The depth and form of the interaction can be explored through a number of proxies of user engagement with the survey:
\begin{itemize}
    \item \textit{Time taken.} In the context of measuring engagement, time is a fundamental metric indicating the extent of spent effort. However, specific time-based metrics must be selected with nuance, given the fact that differences in modalities have secondary temporal effects that are independent from the user's deliberate actions. For example, the time to complete the survey can be inflated by longer response times of the embodied agent. We thus focus on select temporal measures:
    \begin{itemize}
        \item \textit{Responding time.} How much time was spent typing/speaking an answer.
        \item \textit{Transition time.} The length of the transition space between the end of the agent’s question output and the start of the participant’s response. To define the end of the agent’s question, we consider the moment when the embodied agent stops speaking or when the text-based agent concludes its typing animation.
    \end{itemize}
    \item \textit{Response length.} A dedication to provide longer responses is tied to higher engagement.
    \begin{itemize}
         \item \textit{Word count}, as a standard representative of the number of units that convey meaning.
        \item \textit{Character count}, as a more granular complementary metric that could signal effects on word choice and complexity.
    \end{itemize}
    \item \textit{Self-disclosure.} The number of distinct topics, ideas and concepts introduced in the response that relate to personal matters, such as the individual’s personal characteristics, hobbies or previous experiences.
    \item \textit{Sentiment.} Valence of the attitudes that participants expressed in their responses. Notably, this represents sentiment as part of the engagement with the questionnaire rather than attitudes explicitly toward the agent, which is covered by the satisfaction facet. Ordinal values are coded as integers on a range from -1 (negative), through 0 (neutral), to 1 (positive). 
\end{itemize}

Participant satisfaction was assessed traditionally, through explicit feedback in close-ended questions. For capture of varied perspectives, the questions cover following facets:
\begin{itemize}
    \item Naturalness of interaction with the agent.
    \item User experience with the agent.
    \item User experience with the survey.
    \item Preference for agent variant.
\end{itemize}

Coding, having involved simple heuristics and large volumes of captured conversations, was performed through a hybrid method of expert review augmented by an LLM (GPT-4.1). Development of the data labeling prompt was an iterative process, with its initial phase informed by qualitative review of responses. Over several iterations, labeling of the sample was generated by GPT, the results were collaboratively reviewed by researchers and new prompts and strategies were designed to address issues. Explanations of labeling rules and values were made more explicit, data description was added, temperature was lowered to 0.2 to reduce variability and a randomized batch processing strategy was implemented to mitigate the reference group effect. The final prompt, shown in \autoref{app:prompt} \hyperref[app:prompt]{Data labeling prompt}, was launched three times and labels were attributed through majority voting. After a review where 3.6\% of labels were manually corrected, final coding was obtained. The accuracy of automated labels depended on measure, but was highly correlated with reviewed results: Specificity $r_s=.97, p<.001$, Relevance $r_s=.99, p<.001$, Clarity $r_s=.95, p<.001$, Self-disclosure $r_s=.92, p<.001$, Sentiment $r_s=.98, p<.001$.

Scales of the BFI-2-S and CFQ questionnaires were not used as measures. Their relevance to our experiment was as contextual frameworks for applying VAI rather than analytical instruments. Nonetheless, their results are available in the dataset (see \hyperref[sec:data]{Data availability statement}).

\section{Results}
\label{sec:results}

The study took participants similar amount of time to complete regardless of the experimental variant, $U(80) = 941, z = 1.36, p = .17;$ for the embodied agent $M$ = 36.59 minutes, $SD$ = 11.16, for the text-based agent $M$ = 33.50 minutes, $SD$ = 12.69. The three-minute (9.22\%) time overhead between the means can be attributed to the configuration steps and slower response of the ECA. 

In ECA conversations, Heygen determined the end of the participants’ responses correctly in most cases. Only 2 responses out of 380 were completed through the popup window after the conversation. A single participant mentioned this: “No, but I notice that sometimes Judy has cut me off when I haven't finished speaking. Maybe my vocalising (mm or hmm) is too quiet and she thinks I've finished?”. Thirty-one other meaningful responses (excluding “no”, “nothing”, etc.) occurred in the post-conversation popup window only because GPT-4o-mini sometimes ignored the instruction of the Wrapping module and asked another question when the participant could no longer reply using voice.

The most popular ECA models among those shown in \autoref{fig:avatar-appearances} were June (14), Ann (9) and Silas (8). Preferences for models could be attributed to the influence of similarity-attractiveness and social similarity theories within the recruited sample \citep{qiu2010}, or to other ambiguous factors and stereotypes that can influence the traits attributed to the models \citep{aumuller2024}. Only 22.5\% of participants (7 embodied, 11 text-based) interacted with the optional debriefing section of the survey for $M$ = 1.03 minutes ($SD$ = 1.26).

\subsection{Response quality (RQ1)}
\textit{RQ1: Does facilitation of surveys with an embodied conversational agent (ECA) yield different quality of responses to a text-based agent?}

\begin{figure}[!ht]
    \centering
    \includegraphics[width=0.6\linewidth]{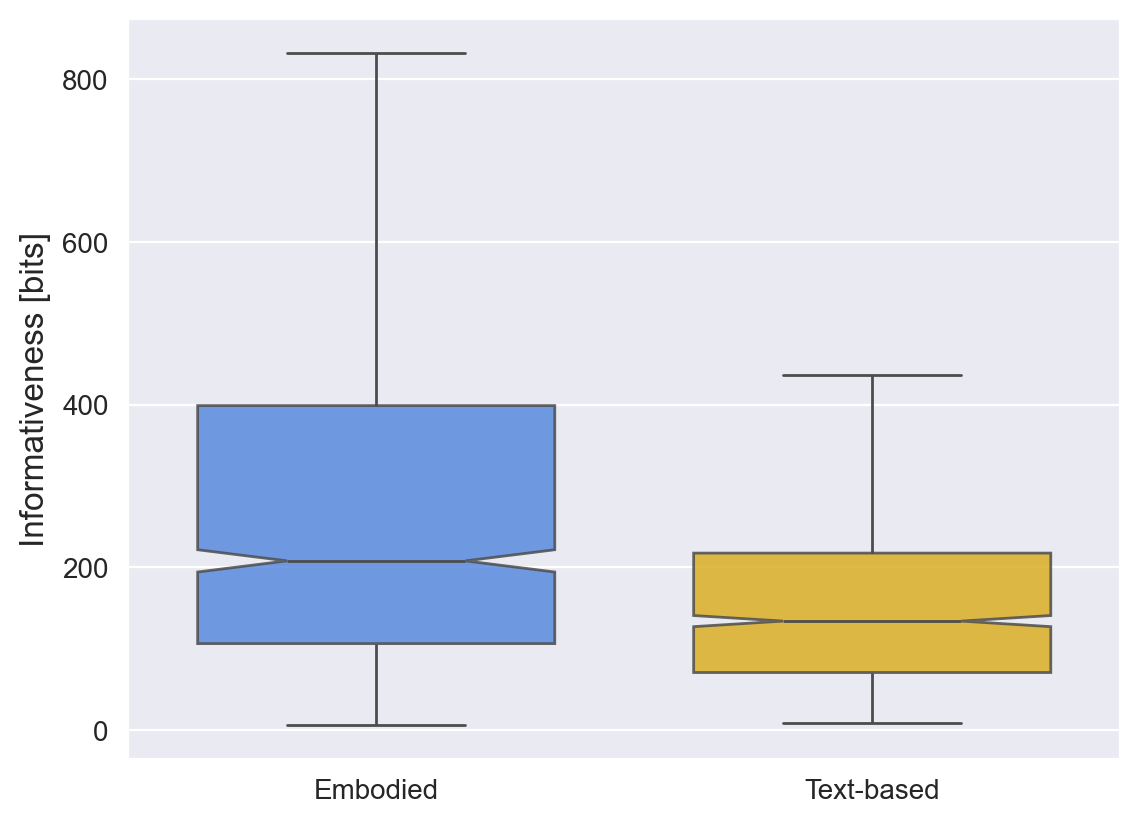}
    \caption{Box plots of informativeness, indicating that participants provided more information to the embodied agent and with a higher degree of variability.}
    \label{fig:informativeness}
\end{figure}

\textit{Informativeness.} Participants generally provided more information to the embodied agent ($M = 285.41, SD = 211.4$) than to the text-based agent ($M = 142.23, SD = 66.49$), to an extent that was statistically significant with high confidence, $U(80) = 1164, z = 3.50, p < .001, r = .39$. Furthermore, the distribution of informativeness in responses in dialogue with embodied agents was significantly more varied ($F(1, 80) = 13.69, p < .001$). The positive skew shown in \autoref{fig:informativeness} signals that the more informative responses tended to expand freely and elaborate their meaning in great detail.

\textit{Specificity.} There was no significant difference between the embodied agent ($M = 1.32, SD = 0.36$) and text-based agent ($M = 1.17, SD = 0.32$), ($U(80) = 969, z = 1.63, p = .10$). Regardless of the agent condition, most responses were sufficiently specific, with a small tendency to further elaborate through detailed examples.

\textit{Relevance.} The overwhelming majority of responses was relevant in both the embodied agent conversations ($M = 1.94, SD = 0.12$) and chats with the text-based agent ($M = 1.99, SD = 0.04$). However, the slightly higher relevance of text-based responses was statistically significant ($U(80) = 555, z = 2.36, p = .003, r = .26$). The difference can be attributed to spontaneity of speech-based responses, a small number of which evaded an explicit answer to the original topic, yet had subtextual implications that remained relevant to the topic.

\textit{Clarity.} While most responses were clear, data indicates strong evidence of small effect on clarity between interactions with the embodied agent ($M = 1.84, SD = 0.18$) and the text-based agent ($M = 1.98, SD = 0.03$). Observation of stochastically lower and more variable clarity of responses to humanlike agents, ($U(80) = 215, z = 5.63, p < .001, r = .63; F(1, 80) = 18.16, p < .001$), could stem from intrinsic differences between writing in a chat and speech. For example, mistakes caused by cognitive load or shifting focus—such as nonstandard verbalizations or word order—could be corrected only in writing. Natural speech also contained unfinished or nonsensical sentences and other artifacts.

This data corroborates the expectation that leveraging an ECA can improve the quality of responses, namely by making participants communicate more information. Although there were some differences in relevance and clarity that could be interpreted as slightly favoring text over speech modality of expression, they were only small and arose from speech as a transient, intuitive and less structured communication channel. Thus spoken responses could be also viewed as more genuine.

\subsection{Engagement (RQ2)}

\textit{RQ2: Does facilitation of surveys with an ECA affect user engagement?}

\begin{figure}[!ht]
    \centering
    \includegraphics[width=0.7\linewidth]{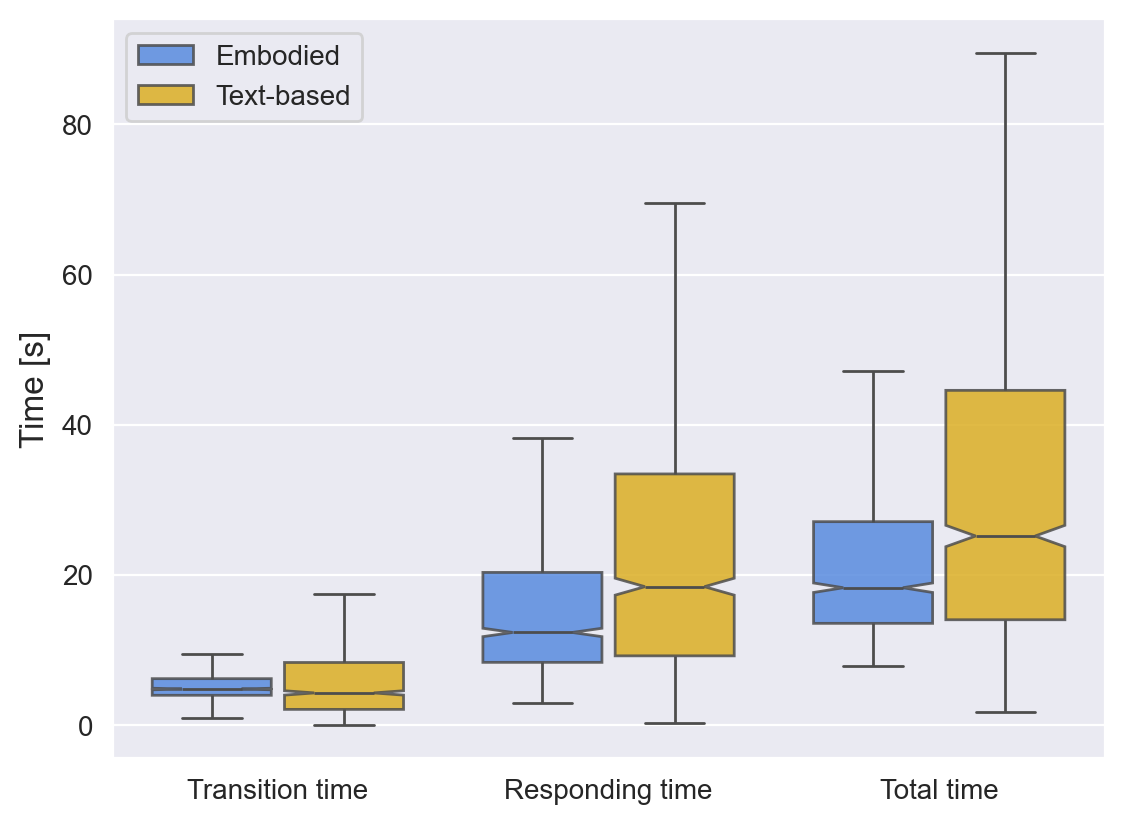}
    \caption{Box plot illustrating the distribution of thinking, answering, and total time for text-based and embodied assistants. Durations tend to be longer for the text-based variants.}
    \label{fig:time}
\end{figure}

\textit{Time taken.} Ranked-based comparison $U(80) = 960, z = 1.54, p = .12$. indicates that Transition time was not significantly higher or lower between embodied ($M$ = 4.99 sec, $SD$ = 1.3 sec) and text-based agent variants ($M$ = 4.73 sec, $SD$ = 3.1 sec). However, it was significantly more variable in the text-based variant ($F(1, 80) = 10.21, p = .002$), indicating its deliberate nature in comparison to speech that was more spontaneous. Responding time was typically longer for the text-based variant ($M$ = 22.02 sec, $SD$ = 13.35 sec) than the embodied variant ($M$ = 15.25 sec, $SD$ = 8.44 sec), as evidenced by statistical significance $U(80) = 531, z = 2.59, p = .010, r = .29$. Due to the higher values of Responding time over Transition time, the statistical difference is predictably reflected in Total time taken $U(80) = 549, z = 2.42, p = .016, r = .27$; participant conversation turns in embodied variant took $M$ = 20.79 sec, $SD$ = 8.48 sec in total and in the text-based variant they took $M$ = 28.8 sec, $SD$ = 15.78 sec. Therefore, in isolation, time differences shown in \autoref{fig:time} could be viewed as signaling lesser engagement in the embodied variant. Due to the intrinsic differences between speech and writing as input modalities however, time could be insufficient for capturing engagement, as corroborated by increased response quality and analysis of following measures.

\begin{figure}[!ht]
    \centering
    \includegraphics[width=0.7\linewidth]{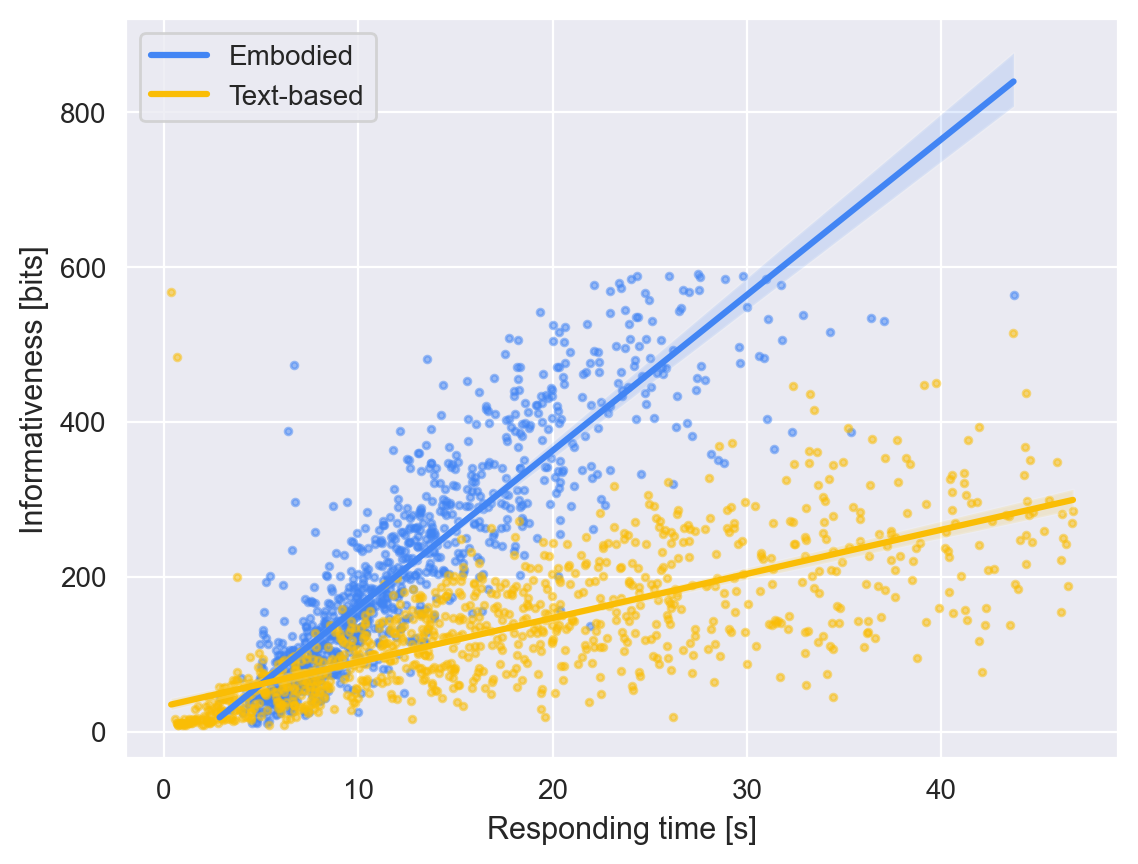}
    \caption{Scatter plot with linear regression illustrating the relationship between Responding time and Informativeness, grouped by conversational agent. Due to noise outliers, IQR is plotted for better visibility of the relationship.}
    \label{fig:scatter}
\end{figure}

Analysis of the relationship between Responding time and Informativeness (the measure of Response quality most affected by embodiment) reveals an influence on communication efficiency. The slope for the embodied agent is significantly steeper, as indicated by the significant interaction term, $\beta = -0.42, SE = 0.094, t(1946) = -4.49, p < .001$, with variables logarithmically transformed and HC3 standard errors used to account for heteroscedasticity of data. As shown in \autoref{fig:scatter}, their comparison indicates more information was communicated faster in interactions with the embodied agent.

\begin{figure}[!ht]
    \centering
    \includegraphics[width=\linewidth]{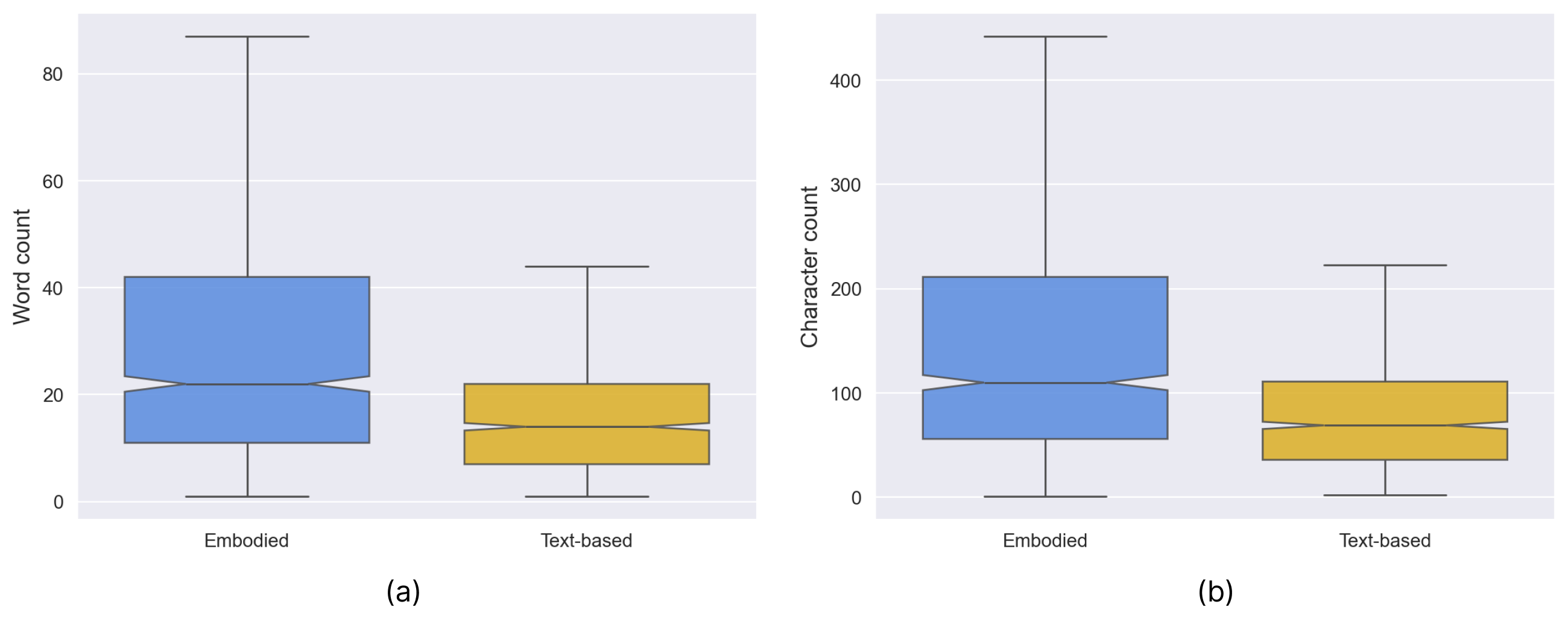}
    \caption{Box plot illustrating the distribution of word counts (a) and character counts (b) for text-based and embodied assistants. Text-based variants’ answers tend to be longer.}
    \label{fig:words}
\end{figure}

\textit{Response length.} On average, participants used more words while conversing with the embodied agent ($M = 29.32, SD = 22.24$) than the text-based baseline ($M = 14.3, SD = 6.89$). This difference, shown in \autoref{fig:words} was statistically significant, $U(80) = 1171, z = 3.57, p < .001, r = .40$, indicating higher verboseness of spoken answers. This is corroborated by character counts being significantly higher in responses to the embodied agent ($M = 149.31, SD = 110$) than in chat ($M = 73.51, SD = 35.12$), $U(80) = 1174.5, z = 3.60, p < .001, r = .40$. While this could be framed as natural—given the findings about word count—it also eliminates the interpretation that the fewer words in chat responses could have been articulated better or more complex.

\begin{figure}[!ht]
    \centering
    \includegraphics[width=\linewidth]{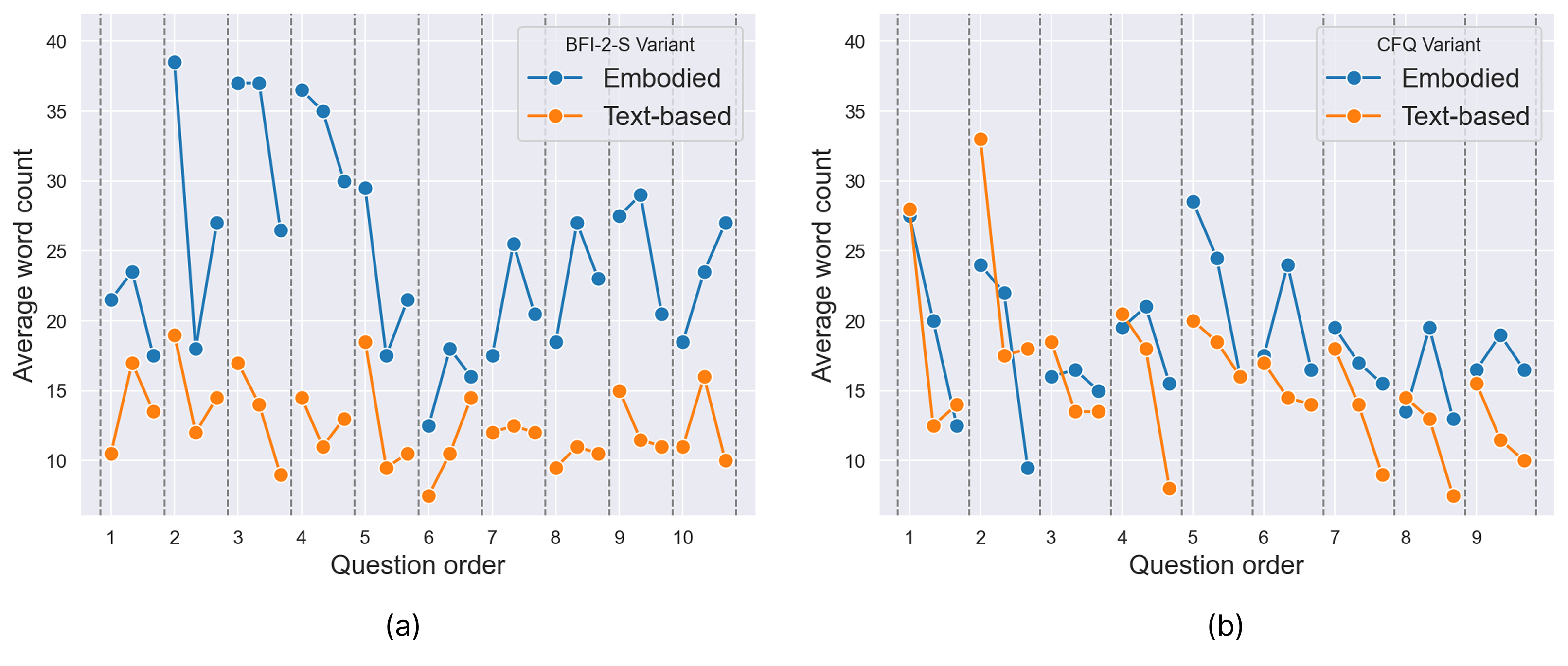}
    \caption{Median word count per follow-up question response, grouped by conversation (follow-up to individual questions). Diagrams correspond to surveys (a) BFI-2-S and (b) CFQ.}
    \label{fig:words-time}
\end{figure}

The temporal distribution of the median response word count declined over time for the text-based agent, while mostly remaining consistently high for the embodied agent. As shown in \autoref{fig:words-time}, BFI-2-S obtained more extensive responses via the embodied agent throughout the survey. The text-based agent in the CFQ survey recorded a downward shift during the final conversations while responses to the ECA remained more consistent in length.

\textit{Self disclosure.} Considering the nature of BFI-2-S and CFQ as self-assessment questionnaires, the number of self-disclosures was high throughout. Embodied agents evoked significantly higher variability in the number of self-disclosures ($F(1, 80) = 9.39, p = .003$). They prompted 17\% (195) responses that contained more than two self-disclosure attributes, versus text-based agents which prompted only 6\% (65) of such responses. However, the difference in average number of self-disclosures between embodied ($M = 1.68, SD = 0.76$) and text-based ($M = 1.32, SD = 0.35$) variants was not statistically significant, $U(80) = 989.5, z = 1.82, p = .069$.

\textit{Sentiment.} There were no statistically significant differences in sentiment between the embodied agent ($M = 0.15, SD = 0.17$) and the text-based agent ($M = 0.17, SD = 0.22$), $U(80) = 798.5, z = 0.01, p = .99$. Prevailing sentiment was neutral with slightly positive tendency (embodied agent 24\% positive and 8\% negative, text-based agent 25\% positive and 7\% negative).

These findings support the expectation that incorporation of an ECA facilitates participant engagement. Social interaction with the avatar through speech encourages participants to speak more than they would write in a chatbot, while also being more time-efficient. Self-disclosures and sentiment are consistent, encouraging their analysis.

\subsection{Satisfaction (RQ3)}

\textit{RQ3: Does facilitation of surveys with an ECA affect user satisfaction?}

Subjective evaluations signaled no significant differences between conversational agents in terms of satisfaction. In the 5-point Likert scales, this manifested as the following:
\begin{itemize}
    \item \textit{Naturalness of interaction with the agent.} $U(79) = 696.5, z = 0.82, p = .37$. Text-based $M = 3.90, SD = 0.93$. Embodied $M = 3.67, SD = 1.08$.
    \item \textit{User experience with the agent.} $U(80) = 683.5, z = 1.12, p = .22$. Text-based $M = 4.40, SD = 0.71$. Embodied $M = 4.12, SD = 0.94$.
    \item \textit{User experience with the survey.} $U(80) = 830, z = 0.29, p = .74$. Text-based $M = 4.45, SD = 0.75$. Embodied $M = 4.5, SD = 0.72$.
\end{itemize}

Preferences for keeping/switching the agent variant differed significantly depending on which agent variant the participants personally experienced, $\chi^2(2, 80) = 16.16, p < .001, V = .45$. Among those who used the baseline text-based agent, 35 preferred to keep using text, 3 would switch to voice and 2 were neutral. The majority preference for text-based interaction could be partially influenced by status-quo bias, linked to the current prevalence of chatbots. Among participants who used the embodied assistant, almost half (18) preferred switching to text, while 13 opted for staying with voice and 9 were neutral. This indicates that interaction swayed approximately half of likely conservative preferences toward embodiment, or at least neutrality. 

Therefore, incorporation of an ECA did not significantly improve user satisfaction. On the other hand, it also did not significantly damage it, in spite of limitations of the VAI prototype (e.g., longer response time), and some participants developed a positive response to the ECA after experiencing it firsthand.

Qualitative analysis was performed to identify issues and appealing factors of the agents. Feedback from participants elucidates their experiences and attitudes. For authenticity, statements below are presented in their original form and grammar, with only glaring typos corrected.

Most participants praised interaction with the ECA as highly natural and helpful for advancing the conversations:
\begin{itemize}
    \item \textit{“It was one of the best conversations I have had with an AI assistant, it was natural, clear and listened.”}
    \item \textit{“I feel like I'm talking with a human.”}
    \item \textit{“It didn’t feel forced and was very pleasant”}
    \item \textit{“The AI has a voice and tone that makes me feel comfortable speaking”}
    \item \textit{“The conversation flowed, I felt like it was really listening to me”}
    \item \textit{“The AI assistant was communicating like it was 100\% real.”}
    \item \textit{“Was really empathetic, and progressed follow-on questions naturally”}
    \item \textit{“The responses were very logical and conversation seemed to flow”}
    \item \textit{“I have used some AI chat bots on websites before, usually customer service bots and they aren't usually anywhere near as intuitive as this AI assistant was.”}
\end{itemize}

Among the minority who did not agree the conversation felt natural, a recurring reservation concerned the agent’s ability to generate meaningful responses. Similar complaints (e.g., repetitiveness, lack of focus on relevant topics for follow-up) were also filed to the text-based agent. This was expected as a consequence of the two agents implementing the same LLM module for conversation logic:
\begin{itemize}
    \item \textit{“She was very responsive and sympathetic to my answers but it was clear that it was AI and not a natural human conversation.”}
    \item \textit{”It was all very generic and didn't feel real”}
    \item \textit{“Though the responses did trigger some interesting thoughts, it was also robotic, and mostly consisted of throwing what I already said back at me with changed tenses.”}
\end{itemize}

More uniquely to the ECA, several participants reported being disrupted by delays between responses, glitches and the speed of speech. Examples:
\begin{itemize}
    \item \textit{“The AI took long pauses before answering and gave very generic statements in reaction to my answers”}
    \item \textit{“The pauses were too long it felt awkward”}
    \item \textit{“The speech was a bit glitchy sometimes and there were sometimes long pauses after I'd finished speaking or the assistant would cut me off mid-sentence. So when it worked, it worked well but because of the glitches it was always very clear that it was AI and not a human.”}
    \item \textit{“The conversation flowed but we had some technical difficulties with freezing and just slow response.”}
\end{itemize}

Additionally, some feedback supplied evidence of the Uncanny Valley effect:
\begin{itemize}
    \item \textit{“She did not understand me, lots of repeating. Weird nodding and movement of the head on a loop”}
    \item \textit{“I felt that some of the AI's responses felt robotic”}
    \item \textit{“Robot voices give me the creeps.”}
    \item \textit{“A slightly strange experience, but if I avoided looking at the screen, the conversation felt fairly natural”}
    \item \textit{“The interaction was interesting and got me to expand on my answers, but the entire concept kind of creeped me out and the knowledge that I wasn't actually talking to a person but inputting my experience as data makes me uneasy.”}
\end{itemize}

When justifying preferences for the agents, participants who preferred interaction with the embodied agent cited intuitiveness, quickness and authentic feeling (e.g., \textit{“Feels more personal, easier and quicker.”}, \textit{“It feels more intimate if voice-based, so more personal”}, \textit{“It helps me pour out more emotions”}). Meanwhile, those who preferred the text-based viewed it as supportive of careful and anxiety-free construction of answers (e.g., \textit{“It's easier for me to read and respond at my own pace, which is made easier by the text based assistant”}, \textit{"I prefer typing as i have anxiety when speaking”}, \textit{“So i will be able to read the text again”}).

\subsection{Survey comparison (RQ4)}

Independent analysis of BFI-2-S and CFQ surveys provides context for more generalizable findings. Although in the majority, indicators behaved similarly across both surveys, there were salient differences that indicate varying effects of embodied agents depending on the research they mediate.

The embodied agent contributed to higher informativeness and word count of participant responses, but it did so significantly only in the BFI-2-S questionnaire (informativeness: $U(40) = 309, z = 2.95, p = .003, r = .47$; word count: $U(40) = 311, z = 3.0, p = .003, r = .47$). CFQ marginally approached significance cutoff for these variables (informativeness: $U(40) = 265, z = 1.76, p = .081$; word count: $U(40) = 268, z = 1.84, p = .067$). Additionally, self-disclosure was stable globally, it was supported more by the humanlike avatar in case of the BFI-2-S. This data indicates that depending on their characteristics, some conversational surveys can benefit from embodied agents more than others, as interpreted further in the discussion.

\section{Discussion}
\label{sec:discussion}

\subsection{Implications}

Based on our findings, virtual agents with human likeness that participants in a survey can talk to present an effective design concept for improving response quality and intensity of participant engagement. By introducing ECAs that simulate social interaction more faithfully than a chatbot, researchers may seek to mitigate hindrances to the quality of responses in unmoderated and remote online research, such as satisficing, cognitive load, lacking motivation, or attention \citep{heerwegh2008}. For user research, this could present opportunities for obtaining detailed qualitative data at high volume, frequency and efficiency.

The higher spontaneity of participants’ spoken responses in contrast to written answers was reflected by occasional reduced clarity and indirect relevance, accompanied by consistently quick transition times. This could result in qualitative data that is more challenging to process and extract insights from. However, the natural quality of spoken responses could be seen as preferable and more genuine. 

To address the slightly lower Relevance of responses given to the embodied agent, the agent’s questions could be also displayed on the screen after their verbalization. This would grant the participant the ability to revisit the question mentally and course-correct at any time during their response. 

With further technological advancements, ECA interviews could potentially be more time-efficient for participants than chatbot surveys. This is supported by spoken responses that were quicker than typing while maintaining higher or similar information value. As a consequence of the ECA's longer response time however, the ECA survey still took longer to complete as a whole. 

By analyzing the conditions of the BFI-2-S and CFQ questionnaires, it could be surmised that characteristics of questions form prerequisites for positive effects on participant responses. While follow-up questions in both variants were open-ended, those in BFI-2-S were broader, pertaining to personality traits that can impact the participant’s life in diverse ways (e.g., experiencing art, music and literature), while CFQ questions were centered more narrowly around specific types of cognitive failures (e.g., distractions and coping with them). Additionally, cognitive failures as part of mental health can be viewed as more sensitive than questions about personality, which could have resulted in slightly more cautious replies, supporting the findings observed by \citet{zhu2025}.

Even though satisfaction ratings were similar between embodiment and chatbot surveys, this is still a positive indicator of the higher human likeness of video-based Heygen avatars. For comparison, prior work by \citet{zhu2025} indicated 3D-modeled avatars (Soul Machines Digital People) as less anthropomorphic than a chatbot. We can draw conclusions about pragmatic and hedonic aspects of user experiences based on thematic analysis of open-ended feedback. Across both variants, criticism linked to conversation contents indicates the critical relevance of verbal communication semantics and pragmatics. It can be argued that without strengthening language modules as a pillar of the conversational agent, the potential of added modalities to create a sense of natural conversation can be limited. As an added benefit, more relevant questions could contribute to more relevant responses.

The positive response of participants who interacted with the ECA—to a degree that some of them would prefer it over a chatbot—is notable. This notion is strengthened by participants attributing human traits to their description of the agent, such as kind, friendly, and sympathetic, or calling the interaction with it pleasant, smooth and coherent. The identified issues with the embodiment (response delays, glitches or repetitive movements evoking the Uncanny Valley effect) can be critical for its user experience \citep{doherty2015}. Addressing them should be a priority to mitigate the biases they might introduce. Measures of satisfaction and self-disclosures might increase for ECAs freed from similar distractions.

\subsection{Limitations and threats to validity}

Application of between-subject design could present internal validity risks, which were addressed through random stratified sampling. Nevertheless, such prevention of carryover bias could be supplanted by other biases (e.g., positivity bias) tied to participants’ lack of interaction with both conversational agent variants. Namely, self-reported ratings of naturalness and user experience are obtained as independent assessments rather than as direct comparisons.

The generalizability of this study is limited by the context of two psychometric questionnaires (BFI-2-S and CFQ) and questions within them selected for conversational agent follow-up. Different types of questions and surveys (e.g., customer experience, competitor analysis surveys) may evoke different attitudes and behaviors when facilitated by conversational agents. The sample may introduce demographic biases, with findings linked to the general population in the UK. High ecological validity was achieved through an experiment conducted in naturalistic conditions typical for online surveys.

Recruitment of participants from an online panel and financial incentives could have introduced a selection bias. Participants were highly motivated, with 0\% attrition. In surveys without the external motivation of financial compensation, internal motivation supported by social interaction could play more critical role in shaping the effect on response quality and engagement. This includes measures that signaled no significant differences during our experiment due to achieving consistently high values, which includes specificity, relevance, clarity, and completion rate.

\subsection{Future work}

Future evolutions of embodied agents in surveys will be contingent on strategies that address delays between conversation turns. During human dialogue, turn-taking offsets are below 0.62s in 75\% of turns and below 1.69s in 95\% \citep{lunsford2016}. For agents incapable of properly responding within these thresholds and holding an uninterrupted real-time conversation, alternative approaches can maintain engagement by obscuring pauses behind conversational conventions \citep{elfleet2024}. As a parallel to humans conversing under cognitive load, natural fillers could be woven into conversations, or turn-taking could be accompanied by gestures such as nodding or touching the hair or the chin \citep{mukawa2014}. These could present more seamless alternatives to the loops of idle head movements and periodic nodding in Heygen, which some participants viewed as robotic.

By validating the core principle of ECA-enhanced surveys, this study lays the groundwork for future research that could expand the concept’s methodological boundaries. This includes designing methods for obtaining better data and improving survey experience, with approaches that could be general or customized for specific survey contents. Aside from tuning virtual avatar and speech modalities, this could include methods for managing and directing the subject of conversations. Research adopting ECAs for various research methods could contribute to the broader vision of developing ECAs for user research, capable of simulating complex moderator conduct, such as monitoring user behavior, speech and expressions, allowing for relevant and timely interceptions.

The methodological limitations of the present study could be addressed by studies with complementary strengths. Alternative experiment designs (e.g., within-subject) and studies focusing on different questionnaires, populations and measures could contribute to the establishment of robust theory. For example, in situations involving sensitive questions, socially desirable responses may be amplified by human likeness \citep{zhu2025}. However, it could also be more challenging to fake convincingly for less sensitive topics due to higher spontaneity, supporting speech-based truthfulness validation for self-reported data \citep{kuric2025lies2}.

\section{Conclusion}
\label{sec:conclusion}

Unmoderated user research makes it easy to collect large amounts of data for understanding users. Yet its lack of genuine social interaction can compromise the quality of its data due to factors like satisficing and perceived lack of immediate accountability. Through the comparison of our novel instrument to the current standard of chatbot-driven surveys, we demonstrate the potential effectiveness of embodied conversational agents (ECAs) for enhancing participant engagement and the quality of responses. By talking to an avatar with a human appearance instead of typing, participants are encouraged to share more in-depth information in a time-efficient manner. Perceptions of agents are divided between agents as personal, intuitive and authentic, and those eliciting Uncanny Valley reactions. For more natural interactions, improvements to ECAs (management of turns, facial animations and gestures, conversation logic) should continue to be explored. While further development and research is required, we conclude that ECAs can represent a positive step for closing the gap between unmoderated and moderated user research aimed at facilitating qualitative insights.

\subsection*{Declaration of competing interests}
The authors declare that they have no known competing financial interests or personal relationships that could have appeared to influence the work reported in this paper.

\subsection*{Funding sources}
This work was supported by the EU NextGenerationEU through the Recovery and Resilience Plan
for Slovakia under the project No. 17I04-04-V05-00029, and co-financed by the Slovak Research and
Development Agency under Contract No. APVV-23-0408. We would like to thank UXtweak j.s.a. for their generous financial contribution to this research, as well as the UXtweak Research team for their technical and expert support.

\subsection*{Data statement}
\label{sec:data}

Supplementary data and materials, including data files containing participants’ responses, transcripts of interactions with the virtual assistant, instructions, interface screenshots, code used for the analysis, along with the source code for the VAI tool are available in the paper repository at \url{https://github.com/moderated-survey-research/embodied-virtual-moderator}.

\appendix

\section{Technical implementation}
\label{app:implementation}

Heygen—the AI video avatar generation tool integrated into VAI—utilizes Whisper\footnote{Whisper speech recognition model: \url{https://platform.openai.com/docs/models/whisper-1}} by OpenAI to transform the user’s speech to text. The Heygen API then processes a prompt through a large language model—GPT-4o-mini\footnote{GPT-4o-mini LLM model: \url{https://platform.openai.com/docs/models\#gpt-4o-mini}}—to generate replies. These are transformed into speech, with lip-syncing and facial expressions in the animated avatar strengthening the embodiment and the realism of the agent's dialogue with the user.

Technical limitations of Heygen shape some of the design aspects of VAI. Only GPT-4o-mini is supported by Heygen. Although other LLM models may achieve better performance in language tasks, GPT-4o-mini is also incorporated in the text-based agent to achieve homogenous conditions where differences relate to embodiment rather than the AI model.

The survey tool is composed of a client-side, built as a React.js application, and a server side that uses Django with a PostgreSQL database and Redis for data caching.

\section{Modularization and logic}
\label{app:modularization}

VAI seeks to approximate the ability of human moderators to flexibly address a variety of problems that arise spontaneously and sometimes simultaneously. To solve complex problems, steps in reasoning can be followed separately to focus on (1) providing relevant replies or (2) asking relevant questions, (3) managing a consistent and high-quality of conversation, even when faced with disruptions, whether malicious or due to low effort, (4) closing conversations in an appropriate manner once the conversation is of sufficient length, and (5) ensuring awareness of the context of previous answers and conversations.

Modularization is used to ensure encapsulation of reasoning steps, and to facilitate the explainability of their outcomes during the prompt design process. Splitting thought processes between multiple modules in LLM-driven solutions can improve their performance \citep{pan2025}. Agent modules differ in their prompts and methods of integration. They include the following:
\begin{itemize}
    \item \textit{Security module.} Participant replies are classified to ensure that they receive an appropriate response. Particularly, this functions as a filter for replies that do not constitute valid information for the conversation. Differentiated classes include direct answers to questions, replies that fit into context, requests for clarification, expressions of confusion, declinations to respond, and replies that are off-topic, nonsensical, or otherwise irrelevant.
    \item \textit{Summary module.} Key descriptive and procedural information about the survey, as well as progress—question answers and conversations—is aggregated into a summary. Refreshed before every upcoming conversation, the summary provides context as part of the input to modules that generate participant-facing replies (Follow-up and Discussion module).
    \item \textit{Discussion module.} During survey instructions and result debriefing, this module generates the agent’s responses in a discussion of the current step (e.g., clarifying and explanations).
    \item \textit{Inquiry module.} In the context of a specific question of the survey, this module is tasked with asking open-ended naturally-flowing follow-up questions. Its purpose is to encourage the participants to express themselves freely and provide more detailed information.
    \item \textit{Wrapping module}. Once the agent has reached the threshold of maximum back-and-forths with the user (10 for Discussion module to encourage exploration, 3 for Inquiry module to mitigate repetition), the wrapping module is tasked with naturally capping off the exchange. In the conversation, this functions as a final output.
\end{itemize}

\begin{figure}[!ht]
    \centering
    \includegraphics[width=0.6\linewidth]{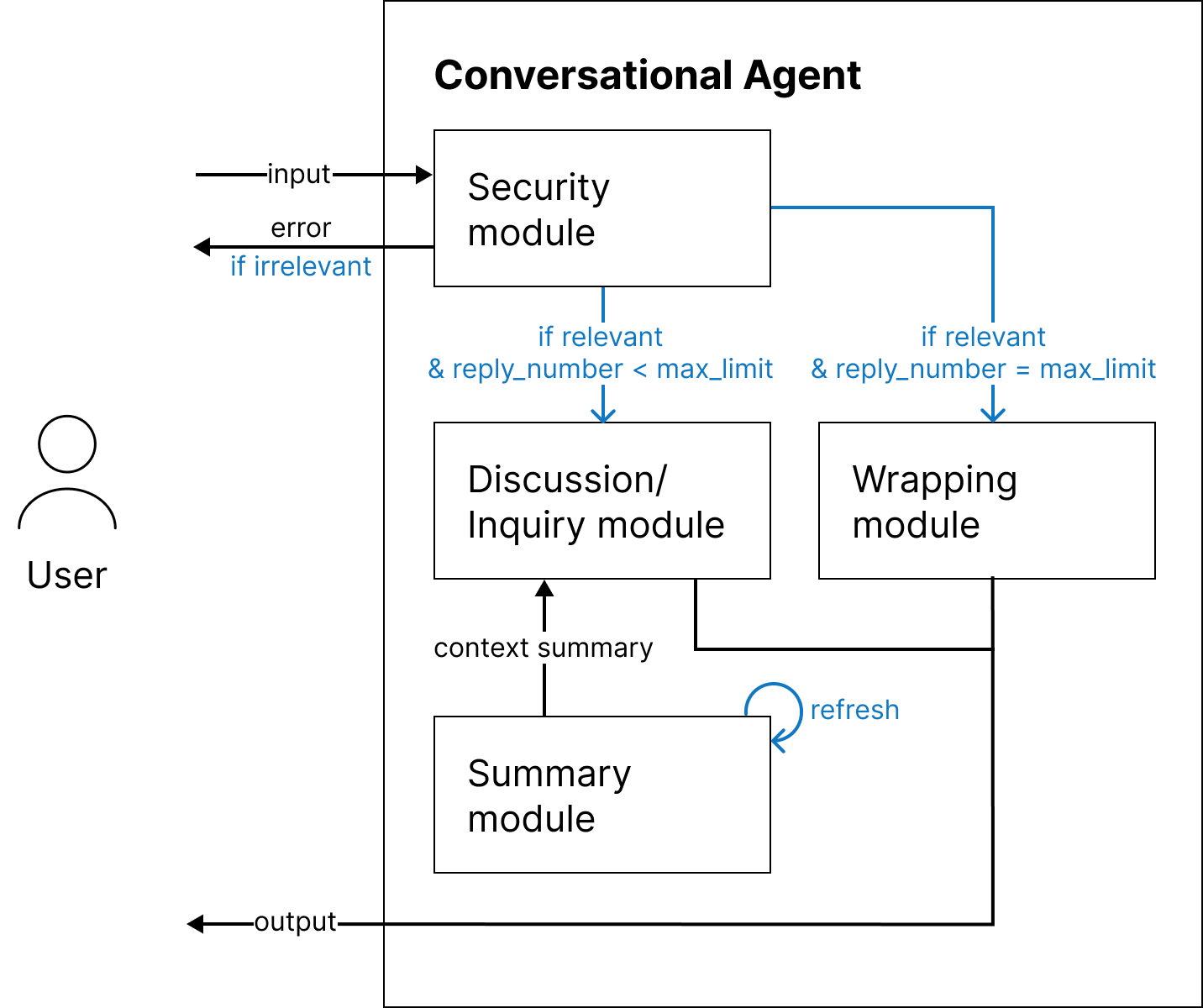}
    \caption{LLM module architecture diagram of VAI conversational agents (text-based and embodied). Displays module interaction and flow of data during user interaction, from user input to system output.}
    \label{fig:modules}
\end{figure}

The conversational model is assembled from modules as shown in \autoref{fig:modules}. In the case of the text-based agent, all modules are implemented through the GPT-4o-mini API. In the case of the ECA, Heygen API has built-in GPT-4o-mini integration, which only allows for a single prompt per reply to be run. This prompt must contain all the information in its body. To ensure an implementation that is as similar as possible, the directives of the ECA’s Security module are integrated into the same prompt as the Discussion and Inquiry modules. The history of the conversation is embedded in the prompt rather than the messages parameter. Neither preliminary testing nor evaluation indicated that the placement of this information causes significant differences in LLMs behavior, such as in responses to irrelevant inputs. Other modules function identically between the variants, with the Summary module running independently to generate contextual input. Heygen is configured to swap to the Wrapping module automatically after reaching the maximum limit on back-and-forths. 

The prototype's module prompts were tuned in an iterative process. Multiple rounds of reviews were conducted by researchers with expertise in user research to reduce individual biases. Before further evaluation, pilot usability testing was also conducted to verify the absence of essential issues that could pollute the results (e.g., security module failing to identify an essential type of answer that the system should account for). Relevant prompts are available in the data repository (see \hyperref[sec:data]{Data availability statement}).

Heygen can sometimes end conversations prematurely (e.g., if participants pause their speech for too long). To assess frequency and scope of unsaid information, the ECA variant displays a popup window after each conversation where participants can write information they wanted to tell but were made unable to so during the conversation.

\section{Data labeling prompt}
\label{app:prompt}
\begin{minipage}{\linewidth}
{\ttfamily \raggedright
Act in the role of a user researcher analyzing conversations between an AI assistant and human participants. Conversations, identified by conversation ID, comprise chronologically ordered questions asked by an AI assistant and answers by the participant, each identified by a message ID. The conversations are from different participants and cover various topics. Assign these measures to each user answer:
\begin{enumerate}
    \item[\texttt{1.}] specificity: how specific is the information given in the answer, on a scale from 0 to 2 (0 = contains general descriptions, 1 = contains specific concepts, 2 = contains specific concepts with detailed examples)
    \item[\texttt{2.}] relevance: how relevant is the answer to the question being asked, rate on a scale from 0 to 2 (0 = irrelevant, 1 = partially relevant, 2 = highly relevant)
    \item[\texttt{3.}] clarity: how clear is the participant’s answer, rate on a scale from 0 to 2 (0 = illegible, 1 = incomplete or partially legible, 2 = clear and well-articulated). Consider semantic rather than syntactic clarity. Typos or joined words (e.g., "dontknow") don’t lower the score if the response is otherwise clear.
    \item[\texttt{4.}] self-disclosure: count of unique personal attributes, topics, concepts, or ideas mentioned by the participant in their answer, such as previous experiences, feelings, hobbies or other personal information (0 or more)
    \item[\texttt{5.}] sentiment: how positive/negative is a participant's answer (1 = positive, -1 = negative, or 0 = neutral) Does the answer express positive/negative attitude toward its subject, or does it describe it neutrally? Do not make assumptions if the sentiment is not sufficiently explicit.
\end{enumerate}

Output a CSV file with six columns: "message\_id" (copied from input), and the five assigned measures. Exclude the original conversation\_id, question and answer columns. Ensure valid CSV formatting.}
\end{minipage}

\bibliography{sources}

\end{document}